\documentclass{aa}

\usepackage{graphicx}
\usepackage{txfonts}
\usepackage{mathtools}
\usepackage{amsmath}
\usepackage{changes}

\begin{document}

   \title{Bolometric night sky temperature and subcooling of telescope structures}
   \titlerunning{Bolometric Night Sky Temperature and Subcooling\ldots}

   \author{
        R. Holzl{\"o}hner\inst{1}
   \and S. Kimeswenger\inst{2,3}
    \and W. Kausch\inst{2}        
    \and S. Noll\inst{4,5}
          }
\authorrunning{Holzl{\"o}hner, Kimeswenger, Kausch and Noll}

   \institute{
European Southern Observatory (ESO), Karl-Schwarzschild-Str.~2, 85748 Garching, Germany
\and
Institut f{\"u}r Astro- und Teilchenphysik, Leopold--Franzens Universit{\"a}t Innsbruck, Technikerstr.~25, 6020 Innsbruck, Austria
\and
Instituto de Astronom{\'i}a, Universidad Cat{\'o}lica del Norte, Av.~Angamos 0610, Antofagasta, Chile
\and Institut für Physik, Universit{\"a}t Augsburg, Universit{\"a}tsstra{\ss}e 1, Augsburg 86159, Germany
\and
Deutsches Fernerkundungsdatenzentrum, Deutsches Zentrum f{\"u}r Luft- und Raumfahrt (DLR), M{\"u}nchener Straße 20, We{\ss}ling-Oberpfaffenhofen 82234, Germany
    }

   \date{Received 06~July~2020; accepted 02~October~2020}

  \abstract
   {The term {\it sky temperature} is used in the literature to refer to different phenomena in different contexts which often leads to confusion. In this work, we study $T_\text{sky}$, the {\it effective bolometric sky temperature} at which a hemispherical black body would radiate the same power onto a flat horizontal structure on the ground as the night sky, integrated over the entire thermal wavelength range of 1--100\;{$\mu$}m.\@ We then analyze the thermal physics of radiative cooling with special focus on telescopes and discuss mitigation strategies.}
  {The quantity $T_\text{sky}$ is useful to quantify the subcooling in telescopes which can deteriorate the image quality by introducing an optical path difference (OPD) and induce thermal stress and mechanical deflections on structures.}
   {We employ the Cerro Paranal Sky Model of the European Southern Observatory to derive a simple formula of $T_\text{sky}$ as a function of atmospheric parameters. The structural subcooling and the induced OPD are then expressed as a function of surface emissivity, sky view factor, local air speed, and structure dimensions.}
   {At Cerro Paranal (2\,600\;m) and Cerro Armazones (3\,060\;m) in the Atacama desert, $T_\text{sky}$ towards the zenith mostly lies 25--50 Kelvin below the ambient temperature near the ground, depending to a great extent on the precipitable water vapor (PWV) column in the atmosphere. The temperature difference can decrease by several Kelvin for higher zenith distances. The subcooling OPD scales linearly to quadratically with the telescope diameter and is inversely proportional to the local air speed near the telescope structure.}
   {
   }

\keywords{Telescopes -- Instrumentation: miscellaneous -- Site testing -- Atmospheric effects - Radiative transfer -- Convection}

   \maketitle
%

\section{Introduction}
Any structure on the ground that is exposed to the night sky experiences a radiation imbalance and therefore typically cools below the ambient air temperature (see \citet{Zhao2019} for an overview). Sky subcooling is strongest at sites with low water vapor column, hence under conditions commonly present at telescope sites (high and dry).\@ As a consequence, the air surrounding a steel truss
for example is cooled, causing a wake of cool air. Light rays passing the cooled air experience an optical path difference which induces a transient wavefront error in a telescope \citep{Wilson2013}.\@  This situation is most severe at small wind speeds and thus became known as the "low-wind effect", and has recently been studied~\citep{Sauvage2015,Milli2018}.\@ A secondary detrimental consequence of subcooling is structural deflection due to thermal contraction, inducing mechanical stress and optical misalignment. Quantifying the effective sky temperature is therefore required when designing advanced ground-based telescopes and analyzing their image quality.

Previous studies \citep{1982SoEn...29..299B,2011HMT....47.1171G,2017Nanop...6...20S,2017SoEn..144...40L,2018JQSRT.209..196L} used wide-angle integrating detectors subtending nearly the whole sky. Moreover, these detectors are tested and calibrated at sites with lower altitudes and higher humidity. To derive the effects of the sky temperature on a telescope structure and the optical elements such as the main mirror, a detailed understanding of sky irradiance as a function of the direction on the sky is required as well. To our knowledge, such a study has not yet been published.

In Section\,\ref{sec:skyPhys}, we discuss the night-sky emissivity and define the effective bolometric sky temperature $T_\text{sky}$.\@ Section\,\ref{sec:thermModel} then sets up a thermal balance model of a structure on the ground exposed to sky radiation and forced convection. Furthermore, we discuss the effects of heat conduction and thermal inertia. Finally, we conclude in Section\,\ref{sec:conc}.


\section{Sky radiance and sky temperature}
\label{sec:skyPhys}

\begin{figure*}[t!]
\sidecaption
\includegraphics[width=120mm]{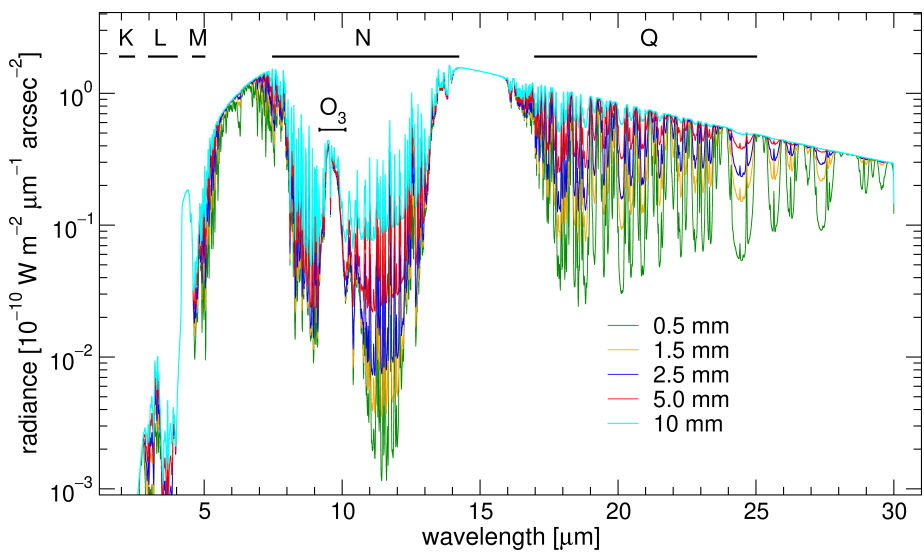}
\caption{Sky radiance $L_{\text{sky}}(\lambda)$ for Cerro Paranal at different PWV conditions typical for observations in the direction of zenith. The astronomical observation 
bands and the main ozone absorption are marked. For further details on species that are not important for us here (e.g., CO$_2$, N$_2$O and CH$_4$) see Fig.\,1 and Table\,1 in \citet{molecfit1}.}
\label{fig_dry_to_wet}
\end{figure*}

The spectrum of the radiance $L_\text{sky}$ for a given area on the sky irradiating a surface strongly depends on wavelength~$\lambda$, zenith distance $\theta,$ and weather conditions. Its total bolometric radiance $E_{\rm BOL}$, defined as the integral of the radiance spectrum over all wavelengths, is dominated by those fractions of the wavelength domain where the atmosphere is completely opaque. In these regions, the mean free path of the light is often a few hundred meters only;  in optically very thick lines it  may even be only a few meters. Thus the main emission is strongly bound to the local temperature near the ground. Those regions where the atmosphere is nearly transparent do not  significantly contribute to the integrated radiance. We use here the common definition of a sky temperature $T_\text{sky}$ as the equivalent temperature of a black-body radiation curve of the same bolometric radiance $E_\text{BOL}$  \citep[see][and references therein]{2017SoEn..144...40L,2018JQSRT.209..196L},
\begin{equation} 
\label{eq:bolometricLuminosity}
E_{\rm BOL} = \int_{0}^{\infty} L_{\rm sky}(\lambda)\, {\rm d}\lambda :=  \frac{\sigma}{\pi}\,\, T_\text{sky}^4,
\end{equation}
where the unit is ${\rm W}\,{\rm m}^{-2}\,{\rm sr}^{-1}$, with the Stefan--Boltzmann constant $\sigma\;=\;5.670367\;\times\;10^{-8}\,{\rm W}\,{\rm m}^{-2}\,{\rm K}^{-4}$. 

\subsection{General trend and climatology of the sky temperature}\label{sec:skytemp_fit}
With our European Southern Observatory (ESO) Sky Model\footnote{https://www.eso.org/observing/etc/skycalc/} \citep{skymodel1,skymodel2} we are able to calculate detailed spectral dependencies and integrate them to derive the bolometric flux and the sky temperature~$T_\text{sky}$.\@ Although this sophisticated model contains a wide variety of components of illumination such as scattered moonlight, zodiacal light, scattered starlight, and atomic and molecular emission lines in the upper atmosphere (airglow), the vast majority of the bolometric flux originates from about $4$ to $7\,\mu$m and longwards of $13.5\,\mu$m and is dominated by the emission of the lower atmosphere.
Our model uses a spectral model of the Earth’s lower atmosphere in local thermal equilibrium calculated by means of the Line-By-Line Radiative Transfer Model~\citep[LBLRTM;][]{lblrtm}.\@ This code is used with the spectral line parameter database {\tt aer}, which is based on HITRAN~\citep{hitran}.\@ Averaged atmospheric profiles for Cerro Paranal describing the chemical composition as a function of height (a combination of the ESO ambient Meteo Monitor, a standard atmosphere for low latitudes\footnote{http://www.atm.ox.ac.uk/RFM/atm/} , and the 3D Global Data Assimilation System (GDAS) model) are also used as input. The GDAS model is provided by the National Oceanic and Atmospheric Administration (NOAA)\footnote{https://www.ready.noaa.gov/gdas1.php} and contains time-dependent profiles of temperature, pressure, and humidity. For the sky model, bimonthly averages over several years were built yielding reliable forecasts for typical conditions at Cerro Paranal. For the transmission and thermal emission of the atmosphere, the current model can be adapted to any site if appropriate GDAS profiles are used. Clear cloudless weather conditions are required with the LBLRTM standard setup.

We show the dependency of the thermal sky radiance as a function of precipitable water vapor (PWV) in Fig.\;\ref{fig_dry_to_wet}.\@ This quantity is equal to the integrated column density across the whole atmosphere, including all layers contributing to the opaqueness, and is given in units of equivalent thickness in millimeters of liquid water. The opaque fraction of the spectral energy distribution resembles a black body radiation curve. While the transparency in the window around $\lambda = 10\,\mu$m is only marginally affected by the H$_2$O molecular bands, it varies strongly in the region from $15$ to $28\,\mu$m (astronomical Q-band) with the water vapor content. Therefore, the weather-dependent variations and uncertainties mainly originate from this wavelength band. There are marginal transparency windows beyond these wavelengths (astronomical mid-infrared Z-band with $28 < \lambda < 40\,\mu$m). As shown by \citet{TAO_PROJECT} , this assumption still holds at sites of even higher altitude than Mauna Kea (e.g., as planned for the $6.5\,$m Tokyo Atacama Telescope project at $5\,640\,$m).\@ Therefore, an approximation with a black body curve is normally sufficient. The ozone absorption in the center of the N-band does not follow the trend of the other opaque regions because of the high altitude of the absorbing molecules.

The sky radiance (expressed as sky temperature as defined above) therefore depends significantly on the~PWV.\@ This circumstance was already found by \citet{1982SoEn...29..299B}, but these latter authors characterized the measured radiation as a function of the relative humidity near the ground, specifically the dew point temperature $T_\text{DP}$ calculated from the humidity. Although radiation from low altitudes prevails, the variation is dominated by semi-transparent regions in the Q-band. Here we see radiation from higher layers having different temperatures and a water vapor content which is not necessarily linked to the relative humidity measured near the ground. 

\begin{figure}[t!]
\centerline{
\includegraphics[width=88mm]{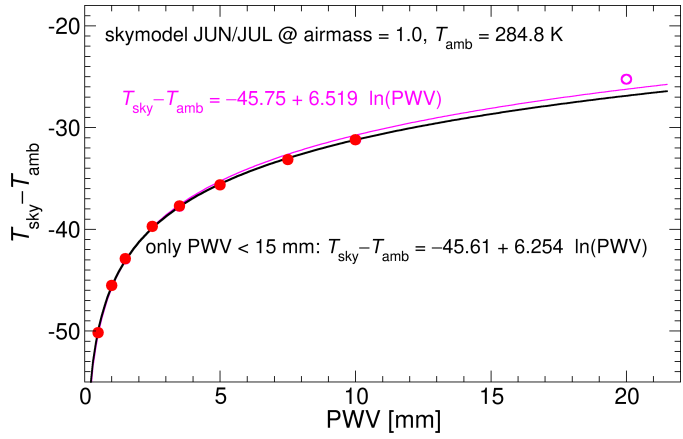}
}
\caption{Correlation of the sky temperature $T_\text{sky}$ and PWV at zenith ($\theta = 0$).\@ Only the range of conditions with $\text{PWV} < 15\,$mm relevant for typical astronomical observations is used for the final fit function.}
\label{fig_pwv_curve}
\end{figure}

\begin{figure}[t!]
\centerline{
\includegraphics[width=88mm]{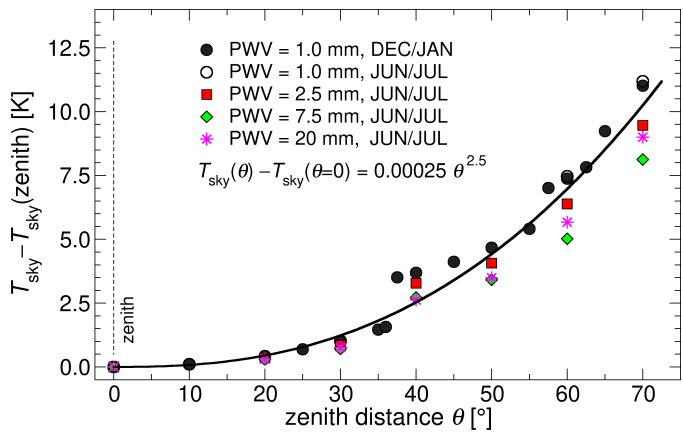}
}
\caption{Correlation of the sky temperature $T_\text{sky}$ and zenith distance at a constant level of~PWV}
\label{fig_airmass}
\end{figure}

We find a phenomenological fit function of the sky temperature with the water vapor content for dry conditions (Fig.\,\ref{fig_pwv_curve}) which is precisely passing the points with deviations of less than $0.1\,$K.\@ In more humid conditions ($\text{PWV} > 15\,$mm) the relation still fits rather well, but has some larger uncertainties, on the order of $2\,$K.\@ Although the fit is based on model data for the southern winter months June and July, it is also valid for the other months of the year. The small differences in the bimonthly mean profiles of pressure and temperature, as well as the relative vertical distribution of water vapor at Cerro Paranal, appear to have a negligible effect on the integrated radiance spectrum.

Moreover, previous studies \citep{1982SoEn...29..299B,2011HMT....47.1171G,2017Nanop...6...20S,2017SoEn..144...40L,2018JQSRT.209..196L} used wide-angle integrating detectors subtending almost the whole sky.
However, in order to calculate a radiation model for a structure, the irradiation of this structure as a function of its {\sl exposed} direction to the sky has to be taken into account (see also Section\,\ref{subsec:Fsky}). This effect is not as important for a structure such as a whole building or a solar cell, as studied by \citet{1982SoEn...29..299B} or \citet{2017SoEn..144...40L} for example. These latter authors therefore needed only integrated values for the entire sky. However, structures of a telescope inside a dome are exposed in a more complex way to different fractions of the sky. Thus, the exact spatial impacts are essential.
When tilting the telescope to higher zenith distances, the radiation is increasingly dominated by low heights. Therefore, we derived the correlation of the irradiance with zenith distance for a wide variety of parameters (Fig.\,\ref{fig_airmass}). A comparison of the summer and winter models shows a very good match of the data points. There are only marginal differences for $\theta=60^\circ$ and $70^\circ$.\@ Therefore, again, a general season-independent fit can be derived.
We find that this correlation does not depend on humidity after accounting for the zenith dependency mentioned above, and a direct empirical fit function with both parameters for good weather conditions (PWV < 15 mm$_{\rm H_2O}$ and no clouds) is found:
\begin{equation}
\label{eq:TSkyFit}
T_\text{sky}(T_{\rm amb},{\rm PWV},\theta) = T_{\rm amb} - 45.75 + 6.52\,\ln({\rm PWV}) + 0.00025\,\theta^{2.5},
\end{equation}
where $T_\text{amb}$ is the ambient temperature near the ground in Kelvin, PWV is in mm$_{\rm H_2O}$, and the zenith distance $\theta$ is measured in degrees.

\subsection{Individual solutions and local estimators}
\label{sec:skytemp_checks}
Our software {\tt molecfit} \citep{molecfit1,molecfit2} was originally designed to correct astronomical spectra for telluric absorption. However, it can also be used to simulate individual thermal emission spectra for a certain date and time at a site. The code automatically derives the GDAS profile of that very day and time from the servers. 

\begin{figure}[t!]
\centerline{
\includegraphics[width=88mm]{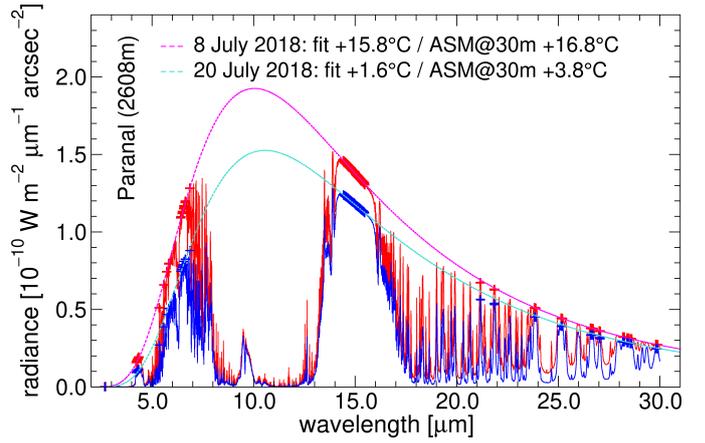}
}
\caption{Sky radiance at zenith on the coldest and the warmest days at Cerro Paranal (2\,600\;m) in July 2018.\@ The fit of a black body is based on the marked points only, defining an upper boundary.}
\label{fig_paranal}
\end{figure}

\begin{figure}[t!]
\centerline{
\includegraphics[width=88mm]{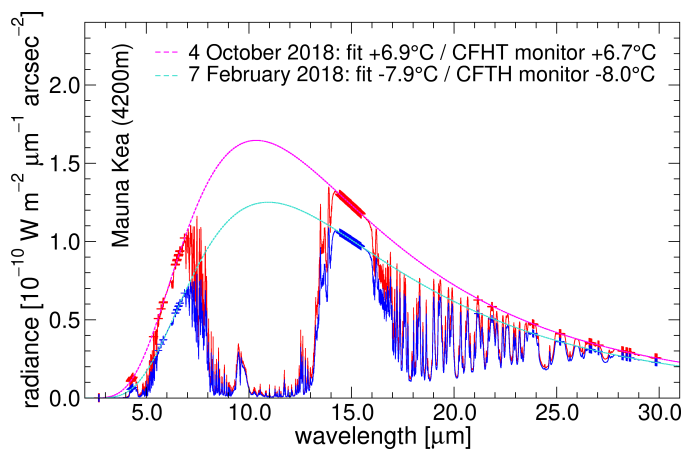}
}
\caption{Same as Fig.~\ref{fig_paranal}, but for Mauna Kea, Hawaii (4\,200\;m)}
\label{fig_hawaii}
\end{figure}

In order to test extreme weather conditions that are not well represented by the sky model, we calculated emission spectra for the warmest and the coldest nights with respect to local midnight at Cerro Paranal in July 2018 (Fig.\,\ref{fig_paranal}).\@ The reference temperatures were measured 30 meters above ground at the ambient monitoring station.\@ They agree well with the temperatures that are derived from a fit of a black body to only those regions in the spectrum where the atmosphere is very opaque. For comparison, we also considered the warmest and coldest nights in 2018 at the high-altitude site at Mauna Kea observatory in Hawaii, as measured at the weather monitor of the Canada-France-Hawaii Telescope (CFHT).\@ This observatory site has significantly lower air temperatures due to its high altitude of 4\,200\;m (Fig.\,\ref{fig_hawaii}).\@ Therefore, the relative humidity at this altitude tends to be lower. However, the latter and the dew point temperature calculated from it \citep{dewT} are only loosely correlated with the integrated water vapor column, that is, the scatter is large \citep{otarola_tmt}.\@ Therefore, we have to use PWV (indicating the impact of the most important and highly variable trace gas H$_2$O) to optimally parameterize the atmospheric radiation \citep{otarola_tmt,masha}. 

While we were able to take the PWV measured from the Low Humidity And Temperature PROfiler (L-HATPRO) microwave radiometer at Cerro Paranal \citep[][see Section\,\ref{sec:meteo_methods}]{ker12} for Fig.\,\ref{fig_paranal}, the PWV values for Hawaii are only based on the (less accurate) GDAS profiles for the given nights. The PWV values and the resulting $T_\text{sky}$ for the integrated sky spectrum and using Eq.\,(\ref{eq:TSkyFit}) are provided in Table\,\ref{table:estimators}. The fit formula causes deviations in $T_\text{sky}$ between $-3.6$ and $+4.8$\,K for these extreme conditions, which suggests that its general uncertainty is not larger than a few Kelvin. 

\begin{figure}[t!]
\centerline{
\includegraphics[width=88mm]{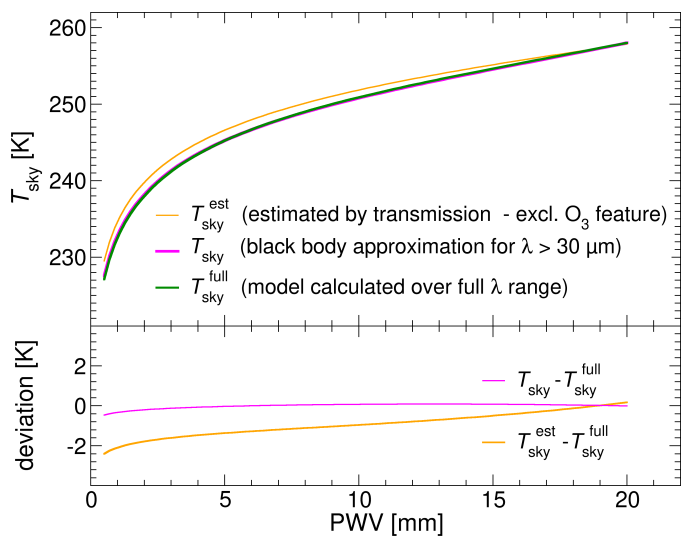}
}
\caption{Comparison of the sky temperature $T_\text{sky}^\text{full}$ derived from the sky model for the full wavelength range, the temperature $T_\text{sky}$ derived from sky models assuming the approximation of pure black body radiation above 30\,$\mu$m, and the estimated $T_\text{sky}^\text{est}$ with the ambient temperature black body radiation and the atmospheric transmission curve only.}
\label{fig_estimate_result}
\end{figure}

In cases where a sky model is not available at a given site, we tested the possibility of estimating the values  of $T_\text{sky}$ from the transmission curve. For the most part, the latter depends on the site altitude and the PWV (which can be obtained by means of GDAS data but might also be derived by various multi-wavelength observations without use of a model), and can therefore be easily tabulated for any site.\@ Multiplying this transmission curve with a black body spectrum only depending on the ambient temperature $T_\text{amb}$ results in the desired estimator $T_\text{sky}^\text{est}$ after integration over all wavelengths (Eq.\,(\ref{eq:bolometricLuminosity})).\@ As indicated in Figs.\,\ref{fig_paranal} and \ref{fig_hawaii}, such an estimate can almost perfectly match the accurate radiative transfer model result and confirms the parameterization in Eq.\,(\ref{eq:TSkyFit}).\@ Only the ozone band from 9.6 to 10.1\,$\mu$m is strongly overestimated because most of the ozone emission originates from a higher layer in the atmosphere with a lower temperature. This kind of emission might be different for very low-altitude sites near populated areas, but the latter is beyond our study. Therefore, the ozone-related wavelength range was excluded from the integration. Other regions with special molecular features do not seem to be dominant enough to significantly contribute to our result. The resulting estimated sky temperature $T_\text{sky}^\text{est}$ in Table\,\ref{table:estimators} does not deviate by more than 2.1\,K from $T_\text{sky}$. For the more typical cases represented by the same parameters as used in Figs.\,\ref{fig_pwv_curve} and \ref{fig_airmass}, $T_\text{sky}^\text{est}$ lies about 1.5\,K above the value of the full model for ${\rm PWV} < 15 {\rm mm}$~(Fig.\,\ref{fig_estimate_result}).

\begin{table}[ht!]
\caption{Test of the extreme conditions at Cerro Paranal and at Mauna Kea: $T_{\rm sky}$ obtained using Eqs.\,(\ref{eq:bolometricLuminosity}) and (\ref{eq:TSkyFit}) as well as $T_{\rm sky}^{\rm est}$ for the individual sky models from Figs.\,\ref{fig_paranal} and \ref{fig_hawaii}.}\label{table:estimators}  
\begin{tabular}{l r c r r r}
\hline\hline
\noalign{\smallskip}
Date & $T_{\rm amb}$ & PWV & $T_{\rm sky}$ & $T_{\rm sky}^{\rm Eq.\,(2)}$ & $T_{\rm sky}^{\rm est}$ \\
& [$^\circ$C] & [mm] & [$^\circ$C] & [$^\circ$C] & [$^\circ$C] \\
\noalign{\smallskip}
\hline
\noalign{\smallskip}
Cerro Paranal:\\
8 Jul 2018 & 16.8 & 1.05\tablefootmark{a} & $-$28.9 & $-$28.6 & $-$26.8 \\
20 Jul 2018 & 3.8 & 0.52\tablefootmark{a} & $-$42.6 & $-$46.2 & $-$44.5 \\
\noalign{\smallskip}
\hline
\noalign{\smallskip}
Mauna Kea:\\
4 Oct 2018 & 6.9 & 2.29\tablefootmark{b} & $-$38.2 & $-$33.4 & $-$36.7 \\
2 Feb 2018 & $-$8.0 & 1.27\tablefootmark{b} & $-$53.5 & $-$52.2 & $-$51.8 \\
\noalign{\smallskip}
\hline
\noalign{\smallskip}
\end{tabular}
\tablefoottext{a}{measurement with L-HATPRO radiometer}\protect{\newline}
\tablefoottext{b}{taken from GDAS model}
\end{table}

The ESO Sky Model yields radiation spectra only up to a wavelength of~30$\,\mu$m. Above this region, pure black body emission is typically assumed in the standard online implementation (SkyCalc).\@ The integration of that emission using Eq.\,(\ref{eq:bolometricLuminosity}) leads to $T_\text{sky}$ used in the rest of the paper. This simplification leads to slight overestimation of the radiation for very dry conditions and high sites. However, a complete model by us using a custom patched version of the radiative transfer model ($T_\text{sky}^{\rm full}$) shows that the correction to the formula given above is marginal for PWV > 1.0\;mm and even below. The discrepancies are of the order of only~0.2\;K~(Fig.\,\ref{fig_estimate_result}).\

\subsection{Ambient meteorological conditions at Cerro Paranal}
\label{sec:meteo}
\subsubsection{Methods and data selection}
\label{sec:meteo_methods}
At Cerro Paranal there are several astronomical site monitoring (ASM) systems installed which permanently record the atmospheric ambient conditions. We investigate data from a period of 4.5 years (2015-07-01T00:00:00 to 2019-12-31T23:59:59) based on the ESO Meteo Monitor, the L-HATPRO radiometer \citep{ker12}, and the Differential Image Motion Monitor (DIMM).\@ An overview of these data is given in \cite{asm}.\@ From the Meteo Monitor we extracted the wind direction, wind speed, and the ambient temperature $T_\text{amb}$ measured at a height of 30 m (as in Section\,\ref{sec:skytemp_checks}) as the telescope structure is more exposed to this part of the atmosphere than towards the ground (c.f. Section\,\ref{sec:skytemp_fit}).\@

The L-HATPRO radiometer database provides integrated PWV measurements at zenith ($\theta=0^\circ$) and measurements of the infrared sky temperature $T_\text{IR}$ achieved by a specific infrared radiometer mounted at the L-HATPRO, which operates in the $9.6$ to $11.5\,\mu$m band. As mentioned in Section\,\ref{sec:skytemp_fit}, the Earth's atmosphere is widely transparent in this wavelength range. Therefore, $T_\text{IR}$ measurements are very sensitive to (comparably warm) clouds at various heights. In particular, ice clouds (cirrus) can be detected, which is not possible with the microwave radiometer.
The DIMM incorporates data on the seeing conditions and the sky transparency, the latter being measured by the relative root mean square (rms) of flux variations of reference stars along the line of sight during nighttime (normalized by the average flux). 

We merged the data points of these three databases via their time stamps by finding the closest Meteo Monitor and DIMM neighbor to each individual radiometer measurement within $\Delta t\leq 300$\,s. By skipping those measurements with $\Delta t>300$\,s, we automatically restrict our merged data set to nighttime because the DIMM only provides data taken between dusk and dawn. We also omitted data points with invalid values and those with PWV$\,\leq 0.02\,$mm.

The data set should represent typical observing conditions at Cerro Paranal, that is, omitting poor weather conditions, which is important because the subcooling of the telescope structure (Section\,\ref{sec:thermModel}) is strongest in the best (=\,photometric) conditions. In addition, the ESO Sky Model does not cover cloudy conditions, and therefore Eq.\,(\ref{eq:TSkyFit}) is not applicable. In order to exclude cloudy periods, we therefore apply a DIMM rms cut of 2\% (which is consistent with the criterion for possible clouds of the online ASM tool at ESO).\@ This cut reduces our sample by only about 10\%;  good photometric conditions prevail at Cerro Paranal. In total we select 276\,958 data points.

\subsubsection{Statistics on the meteorological data}
\label{sec:meteo_results}
\begin{figure}[t!]
\centerline{
\includegraphics[width=88mm]{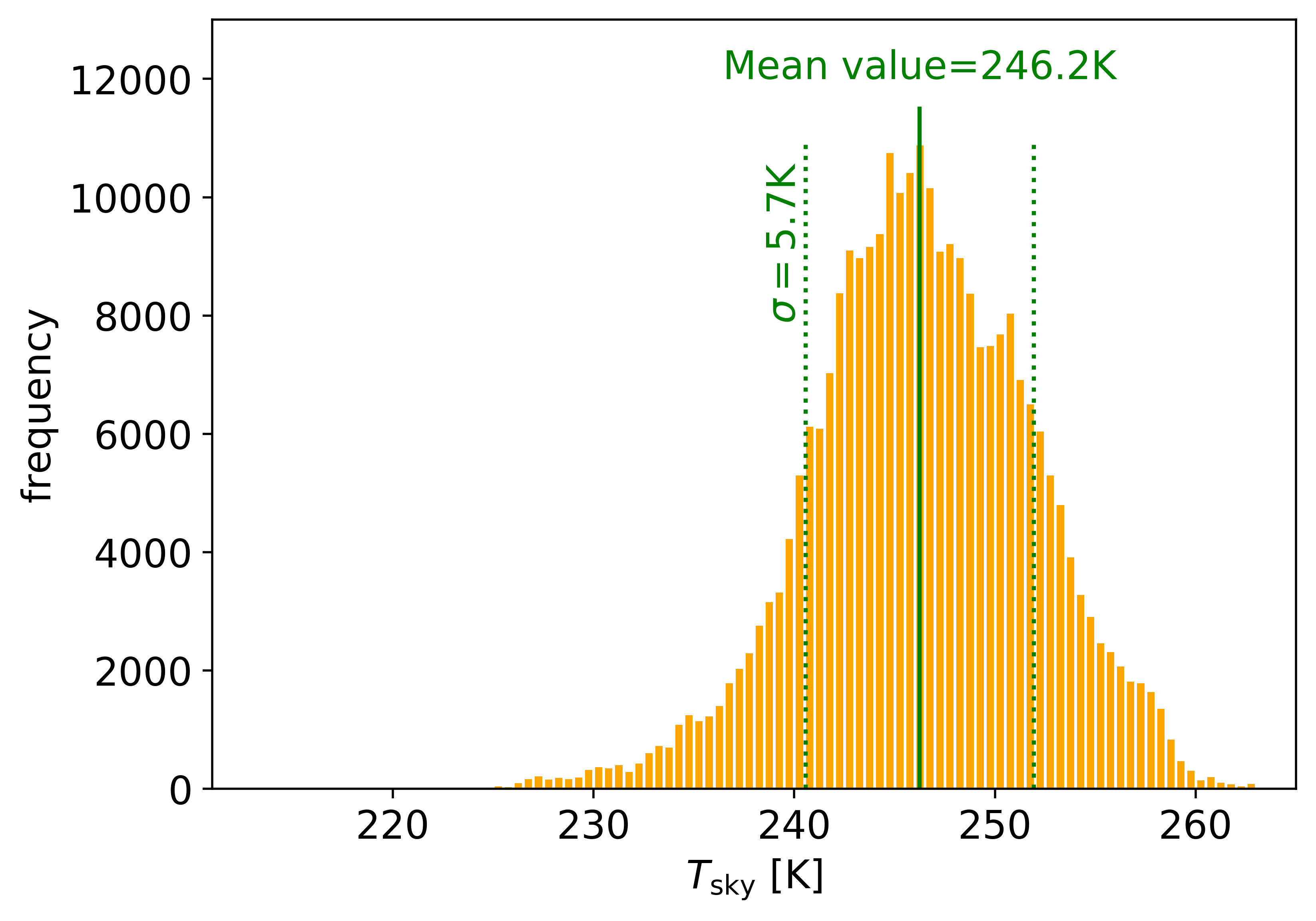}
}
\caption{Histogram of the calculated $T_\textrm{sky}$ values; its mean value is $246.2$\,K, the standard deviation $\sigma=5.7$\,K.}
\label{fig_tsky_hist}
\end{figure}

Investigations of the overall yearly variability reveal a strong trend towards higher PWV values in the southern summer months with an annual median value of PWV$\approx 2.47$\,mm (10\% percentile: $1.06\,$mm; 25\% percentile: 1.56\,mm; 75\% percentile: 4.08\,mm; 90\% percentile: $6.39\,$mm).\@ In addition, weather situations with winds coming from a $90^\circ$ cone centered on the north are the most frequent ones (56\% of the overall time), with a tendency towards higher wind speeds, higher PWV values, and more clouds. As our data selection criteria exclude most of the poor conditions, the sensitivity of our investigations with respect to the wind direction is minimized.
Our empirical equation for $T_\text{sky}$ (Eq.~\ref{eq:TSkyFit}) depends on PWV and the ambient temperature $T_\text{amb}$.\@ We assume these quantities to be independent. Our data set can be used to test this assumption.
\begin{figure}[t!]
\centerline{
\includegraphics[width=88mm]{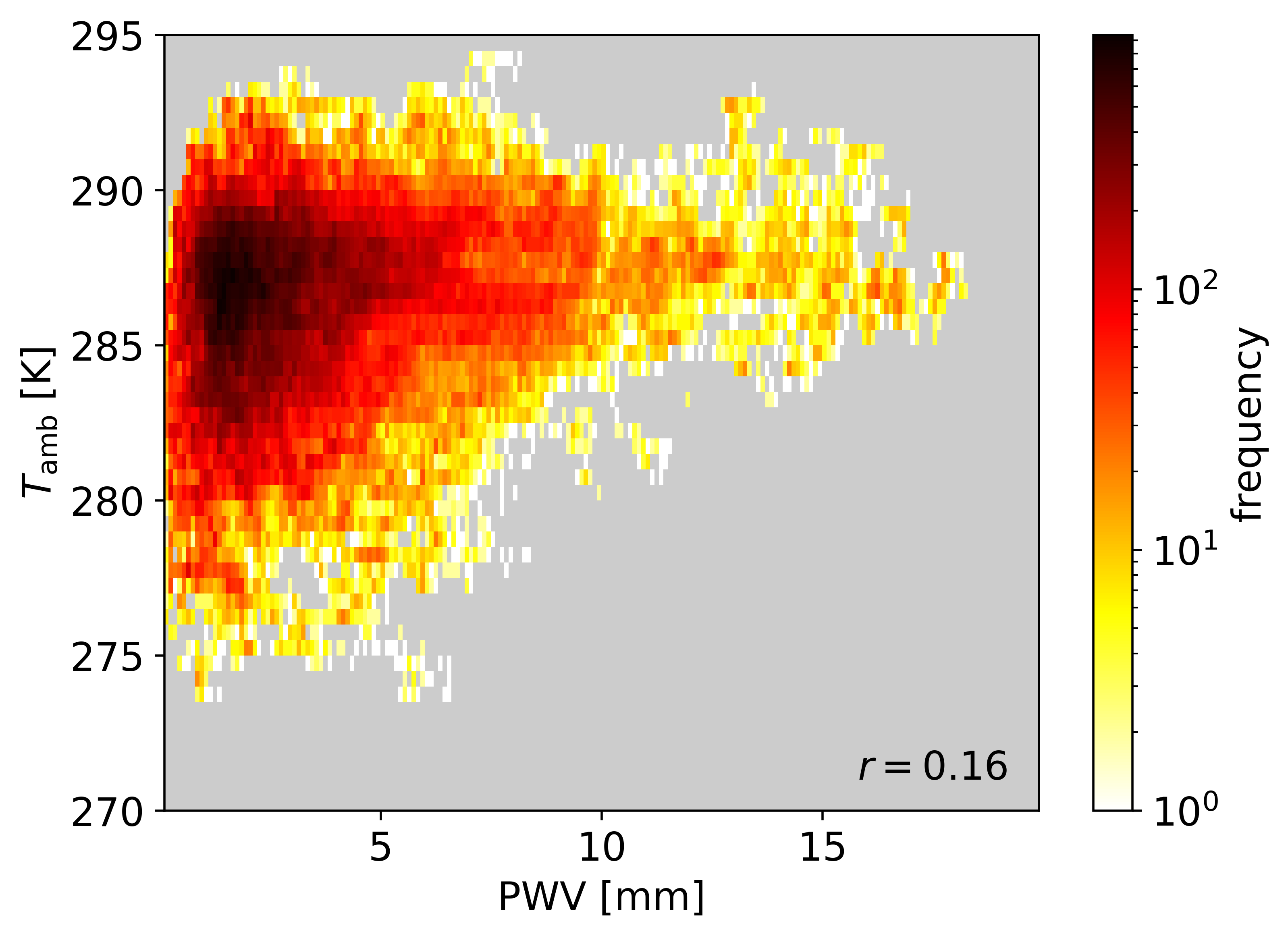}
}
\caption{Two-dimensional histogram of the ambient temperature $T_\textrm{amb}$ as a function of PWV for clear observing conditions (see text for more details).\@ The bin widths of the histogram are $0.1\,$mm in PWV and $0.5$\,K in $T_\textrm{amb}$.\@ The resulting Pearson coefficient $r$ is indicated (see text for more details).}
\label{fig_tamb_vs_pwv}
\end{figure}
Figure\,\ref{fig_tamb_vs_pwv} does not show a clear correlation. This finding can be tested by the Pearson coefficient $r$, which indicates whether or not a data set shows a linear correlation ($r=\pm1$ means perfect linear correlation, $r=0$ no linear correlation).\@ As the absolute value is small in our case ($r=0.16$), we conclude that our parameterization of $T_\text{sky}$ is justified. The distribution of $T_\text{sky}$ given in Fig.\,\ref{fig_tsky_hist} reveals a median value of $T_\text{sky} = 246.2$K and a standard deviation of~$\sigma=5.7$\,K for the selected clear observing conditions.

\begin{figure}[t!]
\centerline{
\includegraphics[width=88mm]{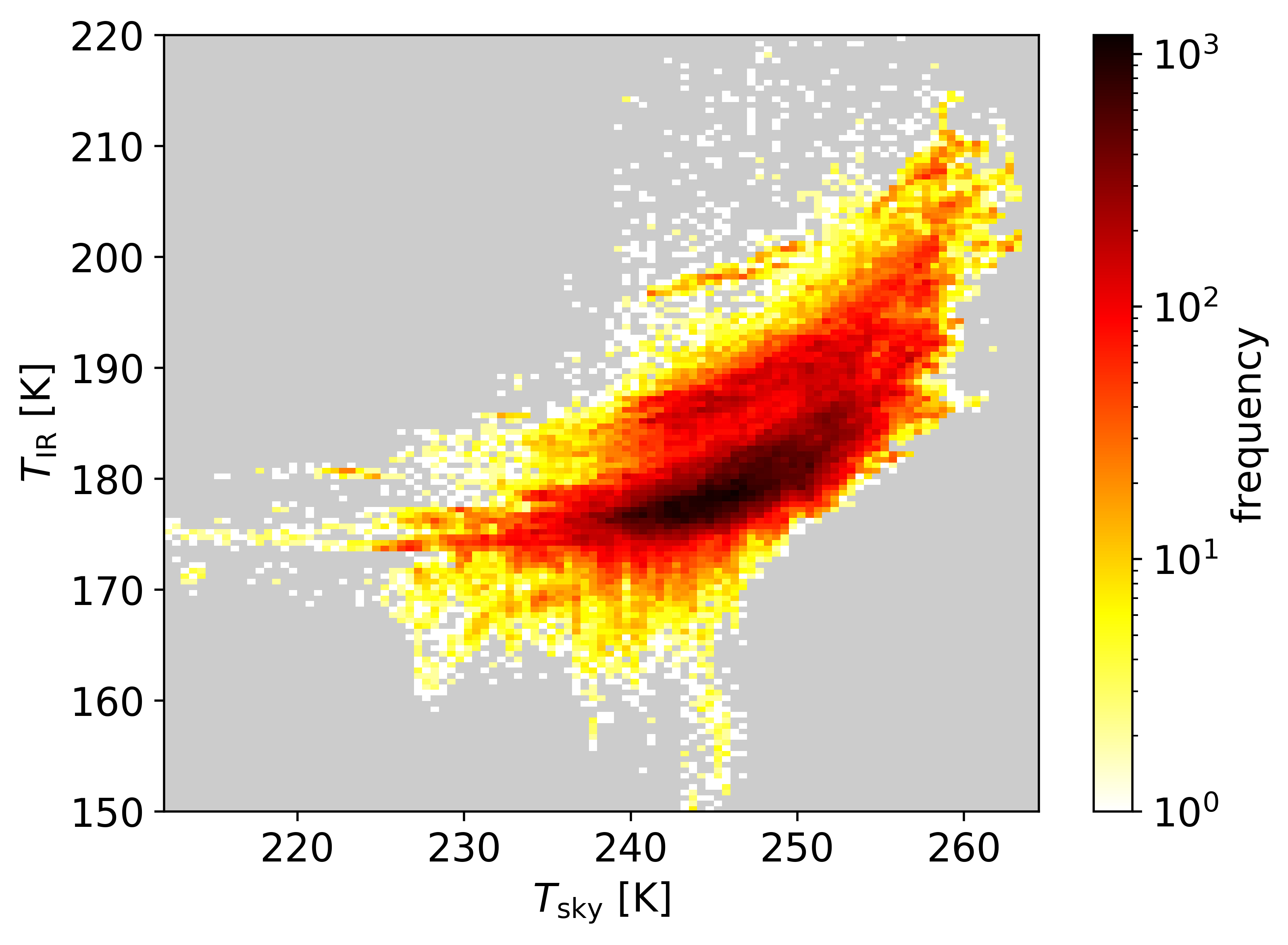}
}
\caption{$T_\textrm{sky}$ vs. $T_\textrm{IR}$ at $\theta=0^\circ$ as two-dimensional histogram (bin width of 0.5\,K) for clear observing conditions. }
\label{fig_tsky_vs_tir}
\end{figure}

Figure\,\ref{fig_tsky_vs_tir} shows the infrared temperature $T_\textrm{IR}$ versus the sky temperature $T_\textrm{sky}$ determined with Eq.\,(\ref{eq:TSkyFit}),  which was calculated by means of the $T_\textrm{amb}$ values from the ESO Meteo Monitor at $\theta=0^\circ$. As $T_\textrm{IR}$ has a relatively weak dependency on PWV but is strongly sensitive to even thin cirrus clouds (due to the small wavelength range used for the measurements), the infrared radiometer temperature is related to the sky temperature $T_\textrm{sky}$ only in the case of photometric observing conditions and $T_\textrm{sky}\gtrsim240\,$K. It is therefore not a good estimator for the sky temperature and its effect on the subcooling of telescope structures. In contrast, $T_\textrm{sky}$  derived by Eq.\,(\ref{eq:TSkyFit}) provides a robust estimate for that purpose only based on measurements of the PWV, the ambient temperature, and the zenith distance~$\theta$.

\subsection{Ambient meteorological conditions at Cerro Armazones}
\label{sec:meteo_ELT}
%

In Section\,\ref{sec:thermModel}, we compute numerical quantities using typical environmental conditions at Cerro Armazones, the site of the Extremely Large Telescope (ELT; 3\,060\,m) \citep{ELTEnvCond20156,Otarola2019}.\@ The median nighttime air temperature is $T_\text{amb} = 9.1^\circ\text{C} = 282.25\,\text{K}$ and the median air pressure equals $p=712$\,hPa. Measurements of PWV such as those taken with L-HATPRO at Cerro Paranal (Section\,\ref{sec:meteo_methods}) are not available for Cerro Armazones. Nevertheless, estimates by \citet{otarola_tmt} and \citet{masha} suggest that the median value for clear sky conditions at Cerro Paranal, namely\ $2.47\,$mm (Section\,\ref{sec:meteo_results}), should be relatively similar. Possible systematic deviations of the order of half a millimeter are not critical for our purposes. Applying Eq.\,(\ref{eq:TSkyFit}) for the selected PWV value, we obtain a difference between $T_\text{sky}$ and $T_\text{amb}$ of $-39.9$\,K at zenith and $-37.8$\,K at the typical observing zenith angle $\theta=37^\circ$.\@ With the ELT median $T_\text{amb}$, this results in $T_\text{sky}$ of $242.4\,$K and $244.5\,$K, respectively.

\citet{Otarola2019} presented a cumulative wind statistic for Cerro Armazones. However, that study was focused on radio astronomy and thus included daytime data. Fortunately, the complete  weather data are publicly available\footnote{https://sitedata.tmt.org} and we selected the nighttime wind data between $22{:}30$ and $9{:}00$\,h~UTC for the years 2006--2009.\@ We find a nocturnal median wind speed at 7~meters above the ground of $7.0$\,m/s, 33\% of the time the wind speed is below $5.2\,$m/s, 25\% of the time below $4.3\,$m/s and 15\% of the time below $3.0\,$m/s.

From a related data set covering the years 2004--2008, we find that the median nocturnal temporal gradient of $T_\text{amb}$ is equal to $-0.34\,$K/h when the temperature is falling and $+0.25\,$K/h when it is rising (which is relevant for significantly shorter time periods).\@ The corresponding quartiles are $-0.17$ and $-0.60\,$K/h (falling) versus\ $+0.11$ and $+0.51\,$K/h (rising), respectively.

%
\relax
\section{Thermal modeling of telescope structures}
\label{sec:thermModel}

\subsection{Radiative heat flux}
In this section, we derive the steady-state temperature of a structure that radiatively cools against the night sky and is heated by forced convection of the ambient air. We consider a small surface of area $A$ that has a spectral emissivity of $\epsilon(\lambda,\gamma)$, where $\gamma$ is the angle between the surface normal $\vec{n}$ and the observing direction. The radiative heat flux from the sky that is absorbed by the surface can be written in spherical coordinates (centered at zenith) according to Lambert's Cosine Law~\citep{Pedrotti1993} as
\begin{alignat}{2}
    \dot{Q}_\text{sky} &= A\! \int_0^{\pi/2} \!\!\text{d}\theta\, \sin(\theta)\! \int_0^{2\pi} \!\!\!\text{d}\phi \, \kappa \!\int_0^{\infty} \!\! \text{d}\lambda \,\, L_{\text{sky}}(\lambda,\theta,\phi)\, \epsilon(\lambda,\gamma),\!\!\!\!\!
    \label{eq:absorbedSkyFlux}
    \\
    \kappa &:= {\text{max}}\bigl(\cos(\gamma),0\bigr),
\end{alignat}
where $L_\text{sky}$ is the spectral radiance from the sky (unit $\text{W}\,\text{m}^{-2}\,\text{sr}^{-1}\text{m}^{-1}$), $\gamma=\gamma(\theta,\phi)$ and $\phi$ is the azimuth. Both $\theta$ and $\phi$ are measured in radians, in contrast to the previous section. If the surface is oriented horizontally ($\vec{n}$ points towards zenith), we find $\gamma=\theta$ and $\kappa = \cos(\gamma)$.

Conversely, to calculate the emitted flux $\dot{Q}_\text{surf}$ from the surface to the sky, one simply replaces $L_\text{sky}(\lambda,\theta,\phi)$ in Eq.\,(\ref{eq:absorbedSkyFlux}) by the black body spectral radiance
\begin{equation}
  L_\text{bb}(\lambda,T_{\text{surf}}) = \frac{2\, h_\text{Pl} \, c^2}{\lambda^5} \,\left[\exp\,\left(\frac{h_\text{Pl}\, c}{\lambda\, k_B\, T_{\text{surf}}}\right) - 1  \right]^{-1},
\label{eq:PlanckRadiance}
\end{equation}
where $T_\text{surf}$ is the surface temperature, $h_\text{Pl} = 6.62618\times10^{-34}\,\text{J}\,\text{s}$ is Planck's constant, $k_B = 1.38065\times10^{-23}\,\text{J}/\text{K}$ is Boltzmann's constant, and $c = 2.99792\times10^{8}$~m/s is the speed of light in vacuum.

As shown in Figs.\,\ref{fig_paranal} and \ref{fig_hawaii}, the sky spectrum agrees with a black-body fit spectrum within several wavelength ranges (e.g., near $7\,\mu$m or near $15\,\mu$m).\@ The temperature of this fit spectrum is close to the ambient air temperature at sites at higher altitudes. The radiation imbalance $\dot{Q}_\text{sky} - \dot{Q}_\text{surf}$ is therefore dominated by the wavelength bands where $L_\text{sky}(\lambda) - L_\text{bb}(\lambda)$ has a large magnitude,  specifically in the astronomical N-band ($7\,$--$\,13.5\,\mu$m) and the Q-band ($17\,$--$\,25\,{\mu}$m).\@ The area between the spectra and their respective fits in Figs.\,\ref{fig_paranal} and \ref{fig_hawaii} provides a graphical representation of this relationship.

The wavelength- and angle-resolved emissivity $\epsilon(\lambda,\gamma)$ is often unknown. Working with an average emissivity~$\overline{\epsilon}$ is possible, but it should be noted that its value is defined by the requirement to preserve the value $\dot{Q}_\text{sky}$ of the wavelength integral in Eq.~(\ref{eq:absorbedSkyFlux}).\@ Any emissivity weighted by the full black-body spectrum, as is sometimes provided by commercial measurement devices or listed in specification sheets, is not well suited to quantify subcooling. Instead, a better approximation is to only consider the emissivity around $\lambda = 11\,\mu$m and, if available, around $18\,\mu$m.

\subsection{Sky view factor}
\label{subsec:Fsky}
The relationship between spectral radiance and $T_\text{sky}$ is \citep{Zhao2019}
\begin{alignat}{2}
  \sigma\, \langle T_\text{sky}^{\;4} \rangle &=  \int_0^{\pi/2} \!\!\text{d}\theta\, \sin(\theta)\, \cos(\theta)\! \int_0^{2\pi} \!\!\!\text{d}\phi  \!\int_0^{\infty} \!\! \text{d}\lambda \,\, L_{\text{sky}}(\lambda,\theta,\phi)
  \label{eq:TSky1}\\
  &= \pi\, E_{\text{BOL}}, \nonumber
\end{alignat}
in accordance with Eq.\,(\ref{eq:bolometricLuminosity}), where the angular brackets denote the geometrical averaging over the sky hemisphere.

Surfaces that are located inside buildings, such as a telescopes in a dome with open doors, can typically only see a solid angle $\Omega < 2\pi$ of the sky hemisphere. The \emph{sky view factor} for a point $\vec{q}$ on the surface is defined by
\begin{equation}
  F_\text{sky} = \frac{1}{\pi} \int_0^{\pi/2} \!\!\text{d}\theta\, \sin(\theta)\,\kappa(\theta,\phi)  \int_0^{2\pi} \!\!\text{d}\phi \;   \mathcal{T}(\theta,\phi),
\label{eq:FSky1}
\end{equation}
where the telescope transmission function $\mathcal{T}(\theta,\phi)$ is equal to 1 at directions $\theta,\,\phi$ for which an unobstructed path from $\vec{q}$ to the sky exists and zero if the path is blocked by any warm black body. However, we also allow for partly obscured ray paths like propagation through a windscreen mesh or indirect ray paths to the sky through reflection on a mirror (often the primary mirror M1), or any partially reflective surface like a bare metal for which $0 < \mathcal{T} < 1$.\@ The normalization factor $\pi^{-1}$ ensures that $0 \leq F_\text{sky}\leq 1$.

The sky view factor for a horizontal surface inside a building with a circular aperture centered at $\vec{n}$ that is seen from $\vec{q}$ under the one-sided angle $\theta_a$ is equal to $\sin^2(\theta_a)$. This term assumes the value $1/2$ at $\theta_a = 45^{\circ}$, although the solid angle of this aperture subtends only $29.3$\% of the hemisphere. If, on the other hand, the aperture is not centered at $\vec{n}$, $F_\text{sky}$ will be smaller. This reasoning shows that the orientation of surfaces inside a telescope dome is important, and not simply the distance from the dome slit. The difference is particularly large when comparing the sky view factors for the top side of the spider trusses (with rectangular cross section) or the secondary mirror housing ($F_\text{sky} \approx 0.8\,$--$\,1$) with its side faces where $\vec{n}$ is perpendicular to the telescope axis. Under no circumstances can the side faces see more than half of the sky hemisphere ($F_\text{sky} \leq 1/2$).\@ Furthermore, the zenith angles $\theta$ at which the cosine weighting term $\kappa$ is large are almost perpendicular to the telescope axis; more precisely, these rays may often point near the horizon (the side faces may even see part of the ground outside the telescope building).\@ The average sky temperature is higher in these directions due to the $\theta^{\,2.5}$ dependence; see Eq.~(\ref{eq:TSkyFit}).

\begin{figure}[htb!]
\centerline{
\includegraphics[width=78mm]{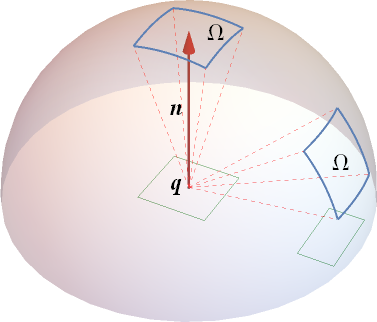}
} 
\caption{Evaluation of sky view factors by projection. Red arrow: Normal vector $\vec{n}$ above the point $\vec{q}$ on a surface. Blue quadrilaterals: Two windows of solid angle $\Omega \approx 0.16\,$sr through which the sky can be seen from $\vec{q}$ (i.e., $\mathcal{T}=1$); the right window is tilted against $\vec{n}$ by $\gamma = 60^\circ$.\@ Green rectangles: Projection of the windows onto the tangential plane with areas $\pi\,F_\text{sky} \approx \Omega$ and $\Omega\times\cos(60^\circ) = \Omega/2$, respectively.}
\label{fig:F_sky_sketch}
\end{figure}
When dealing with noncircular apertures to the sky (e.g., the open door slit of a telescope dome) as sketched in Fig.~\ref{fig:F_sky_sketch}, it can be more convenient to work with Cartesian coordinates~$x,\,y,\,z$.\@ The effect of $\kappa$ in Eq.\,(\ref{eq:FSky1}) is simply to project the area of the sky hemisphere that can be seen from the point $\vec{q}$ onto the unit disk on the tangential surface (perpendicular to~$\vec{n}$).\@ If we let $\vec{\rho} = (x,y,z)$ be a vector of unit length that points from $\vec{q}$ to the sky, and $x,y,z$ are local coordinates, meaning that any vector $(x,y,0)$ lies on the tangential surface and $\vec{n}=(0,0,1)$, we can write
\begin{equation}
  F_\text{sky} = \frac{1}{\pi} \iint_{x^2+y^2 < 1} \!\!\!\!\text{d}x\,\text{d}y \;   \mathcal{T}(x,y).
\label{eq:FSkyCartesian}
\end{equation}
In buildings, the sky windows often have much more complicated shapes than in Fig.~\ref{fig:F_sky_sketch}.\@ One can numerically evaluate $F_\text{sky}$ in this case by describing the edges of the sky windows by a closed set of straight lines. The projected window then forms an (irregular) polygon on the unit disk whose area can be computed easily by the "shoelace formula" \citep{Braden1986}.\@ Dividing this area by $\pi$ yields $F_\text{sky}$.\@ Any window edges whose rays $\vec{\rho}$ would point below the tangential plane, that is, if $\gamma \geq 90^\circ$ and thus $\kappa = 0$, generate projected points that lie on the unit circle in this scheme.

A collection of formulas to calculate $F_\text{sky}$ in special geometries is found in Chapter~10 of~\citet{Lienhard2011}.\

If a structure is exposed to some part of the sky through a mirror, or in general through a series of reflective optics, $F_\text{sky}$ is defined by the ray bundle emanating from $\vec{q}$ that ultimately reaches the sky, hence with~$\mathcal{T} > 0$.\@ In other words, the effective aperture is the image of the real aperture (say, the rim of the dome slit) as seen from~$\vec{q}$.\@ One can compute $F_\text{sky}$ by evaluating the angles $\gamma$ of the ray segments between $\vec{q}$ and the first reflective surface. This method is correct even if the mirrors are powered (i.e., have nonzero curvature) by the principle of {\'e}tendue conservation \citep{Chaves2017} and assuming that the sky is a Lambertian radiator (setting aside the dependence of $T_\text{sky}$ on~$\theta$).\
\subsection{Thermal balance equation}
With the help of Eqs.\,(\ref{eq:absorbedSkyFlux} -- \ref{eq:FSky1}), we can write the radiative heat loss due to sky subcooling as~\citet{Lienhard2011}
\begin{alignat}{2}
    \dot{Q}_\text{sky} - \dot{Q}_\text{surf} &= A\, \overline{\epsilon}\, F_\text{sky}\, \sigma \left(\langle T_\text{sky}^{\;4}\rangle_\mathcal{T} - T_\text{surf}^{\;4} \right) \;\;<\; 0,\\
    T_\text{surf} &= T_\text{amb} + \Delta T_\text{surf},
\end{alignat}
where $T_\text{amb}$ is the ambient air temperature inside the dome on the upwind side of the telescope structure and $\Delta T_\text{surf}$ quantifies the surface subcooling. The expression $\langle T_\text{sky}^{\;4}\rangle_\mathcal{T}$ denotes the average sky temperature with the same transmission $\mathcal{T}$ as in Eqs.\,(\ref{eq:FSky1},\ref{eq:FSkyCartesian}).\@ As discussed in the previous section, this average geometrically depends on the patch of sky seen from $\vec{q}$ and the telescope pointing.

On the other hand, convection heats the surface at a rate of
\begin{equation}
    \dot{Q}_\text{conv} = A\, h\, \left(T_\text{amb} - T_\text{surf} \right) = - A\, h\, \Delta T_\text{surf} \;\;>\; 0,
    \label{eq:convective_heating}
\end{equation}
where $h$ is the interface heat transfer coefficient (in units of $\text{W}\,{\text{K}^{-1}}\,{\text{m}}^{-2}$).\@ In thermal equilibrium and neglecting thermal conduction, we require that
\begin{equation}
    \dot{Q}_\text{sky} - \dot{Q}_\text{surf} + \dot{Q}_\text{conv} = 0,
    \label{eq:Q_balance}
\end{equation}
and therefore the heat fluxes are balanced. Our goal is to quantify the steady-state plate subcooling $\Delta T_\text{surf} < 0$ that satisfies
\begin{alignat}{2}
    \Delta T_\text{surf} &= \sigma\, \omega\, \left(\langle T_\text{sky}^{\;4}\rangle_\mathcal{T} -\Bigl(T_\text{amb} + \Delta T_\text{surf} \Bigr)^4 \right),
    \label{eq:DeltaTSurf1}
    \\
    \omega &:= \frac{\overline{\epsilon}\, F_\text{sky}}{h}.  \nonumber
\end{alignat}
Expression~(\ref{eq:DeltaTSurf1}) is a quartic equation with two real and two complex roots. One of the two real roots lies below $-1300\,$K, and so there is only one meaningful solution. We calculated the latter in closed form, but the expression is rather complicated. Instead, we carry out a Taylor expansion to first order in $\Delta T_\text{surf}$ about $\Delta T_\text{surf} = \delta T$.\@ Solving for $\Delta T_\text{surf}$, we obtain
\begin{alignat}{2}
    \Delta T_\text{surf} &= \frac{\sigma\, \omega}{1 + 4\,T_0^{\;3}\,\sigma\,\omega} \left(  \langle T_\text{sky}^{\;4}\rangle_\mathcal{T} - T_0^{\;3} \bigl(T_\text{amb} - 3\,\delta T\bigr) \right),
    \label{eq:DeltaTSurf2}
    \\
    T_0 &:= T_\text{amb} + \delta T. \nonumber
\end{alignat}
In the remainder of this paper, we choose $\delta T = -3$\,K so that we cover the typical range of subcooling in telescopes of about $-6\,\text{K} \leq \Delta T_\text{surf} \leq 0$ with good accuracy.

Following \citet{Lienhard2011}, we define the radiation heat transfer coefficient 
\begin{equation}
    h_\text{rad} := 4\,T_0^{\;3}\,\sigma\,\overline{\epsilon}\, F_\text{sky} \approx  4.9\,\text{W}\,\text{K}^{-1}\,\text{m}^{-2} \;\;\times\;\; \overline{\epsilon}\, F_\text{sky},  \label{eq:h_rad}
\end{equation}
and, rewriting Eq.\,(\ref{eq:DeltaTSurf2}), finally express $\Delta T_\text{surf}$ as 
\begin{alignat}{2}
    \Delta T_\text{surf} &= \frac{\eta}{1 + \eta}\,\Delta T_D,  \label{eq:DeltaTSurf3}
    \\
    \Delta T_D &:= \frac{\langle T_\text{sky}^{\;4}\rangle_\mathcal{T}}{4\,T_0^{\;3}} - \frac{T_0 - 4\,\delta T}{4} \approx 0.89\,\Delta T_\text{sky},
    \\
    \eta &:= \frac{h_\text{rad}}{h},
\end{alignat}
where the number $\eta$ quantifies the subcooling efficiency. We thus find that $\Delta T_\text{surf}$ scales linearly with $\eta$ as long as $\eta \ll 1$.\@ We highlight that $\Delta T_\text{surf} \propto 1/h$ in this unsaturated regime, which applies to most cases of interest in telescope subcooling. At ELT median conditions (Section~\ref{sec:meteo_ELT}), $\eta$ is of the order of $0.1$ and the subcooling is driven by the temperature offset $\Delta T_D$ which equals $-33.9\,$K for $\theta=37^\circ$.\@ In the opposite limit of $\eta\!\rightarrow\!\infty$, we obtain $\Delta T_\text{surf}\!\rightarrow\!\Delta T_\text{sky} := T_\text{sky} - T_\text{amb}$.\@ In practice, this extreme can probably only be reached on a black surface facing the sky inside a vacuum vessel with a window that transmits infrared radiation.

\subsection{Convective heat transfer}
\label{subsec:convection}
The thermal coupling of a structure to the ambient air through forced convection gives rise to a boundary flow problem~\citep{MartinezBoundaryFlow}.\@ The convective heat transfer coefficient $h$ depends on the air speed $v$ near the surface (which is often much lower than the wind speed outside the dome), the air pressure $p,$ and, to a great extent, on the shape of the structure. The structures closest to or even inside the telescope beam are usually steel trusses supporting the secondary and possibly the tertiary or quaternary mirrors. The simplest truss shape is cylindrical with diameter~$d$.\@ An approximate expression for $h_\text{cyl}$, modeling the net heat exchange of airflow perpendicular to a cylinder, was derived by \citet{Churchill+Bernstein1977}.\@ Their original formula is expressed in terms of the Reynolds, Nusselt, and Prandtl numbers \citep{MartinezConvection} and we checked that the values of these numbers for typical telescope structures are within the validity range of the formula. We do not reproduce the general formula here but instead simplify it to
\begin{alignat}{2}
    h_\text{cyl} &= c_1 \frac{\sqrt{\beta}}{s}  \left(1 + c_2\,\beta^{5/8}\right)^{4/5}, \label{eq:h_cyl}\\
    \beta &:= m\, s\, p\, v,  \label{eq:h_cyl}   \nonumber
\end{alignat}
where $c_1 = 0.0179\,\text{W}\,\text{K}^{-1}\,\text{m}^{-1}$, $c_2 = 1.5\times 10^{-4}$ for airflow at $T_\text{amb} = 282\,$K, and $s = \pi\,d$ is the cylinder circumference. We introduced the Reynolds number scaling quantity $m,$ which equals $1.0\,\text{m}\,\text{W}^{-1}$ if the incoming airflow is perfectly laminar and the cylinder has a smooth surface, and $m \approx 1.5\,\text{m}\,\text{W}^{-1}$ for partly turbulent inflow. However, if a precision of more than about 20\% in $h_\text{cyl}$ is demanded, one has to resort to computational fluid dynamics (CFD) simulations.

At $v=7.6$\,m/s and $s=0.6\,\pi = 1.88$\,m, one reaches $c_2\,\beta^{5/8} = 1$.\@ For much lower air speeds (more precisely, if $\beta \ll 1.3\times10^6$), the scaling law $h_\text{cyl} \propto (m\,p\,v / s)^{1/2}$ applies. We note that, although $v=7.6$\,m/s is close to the median wind speed on Cerro Armazones (cf. Section~\ref{sec:meteo_ELT}), the air speed inside the dome is often much lower. This holds in particular for structures below the M2 level, and therefore this regime is normally valid.

In the opposite limit of large $\beta$, for example for high air speed and/or large structures, $h_\text{cyl} \propto \beta/s = m\,p\,v$.\@ The Reynolds number then becomes so large that the boundary layer is turbulent throughout, and thus the structure size is no longer relevant for~$h_\text{cyl}$.\@ An example for this limit is subcooling of the outer dome cladding, where $v$ approaches the free air wind speed and the interface length $s$ can amount to tens of meters for streamlines that follow the dome surface.

\begin{figure}[htb!]
\centerline{
\includegraphics[width=88mm]{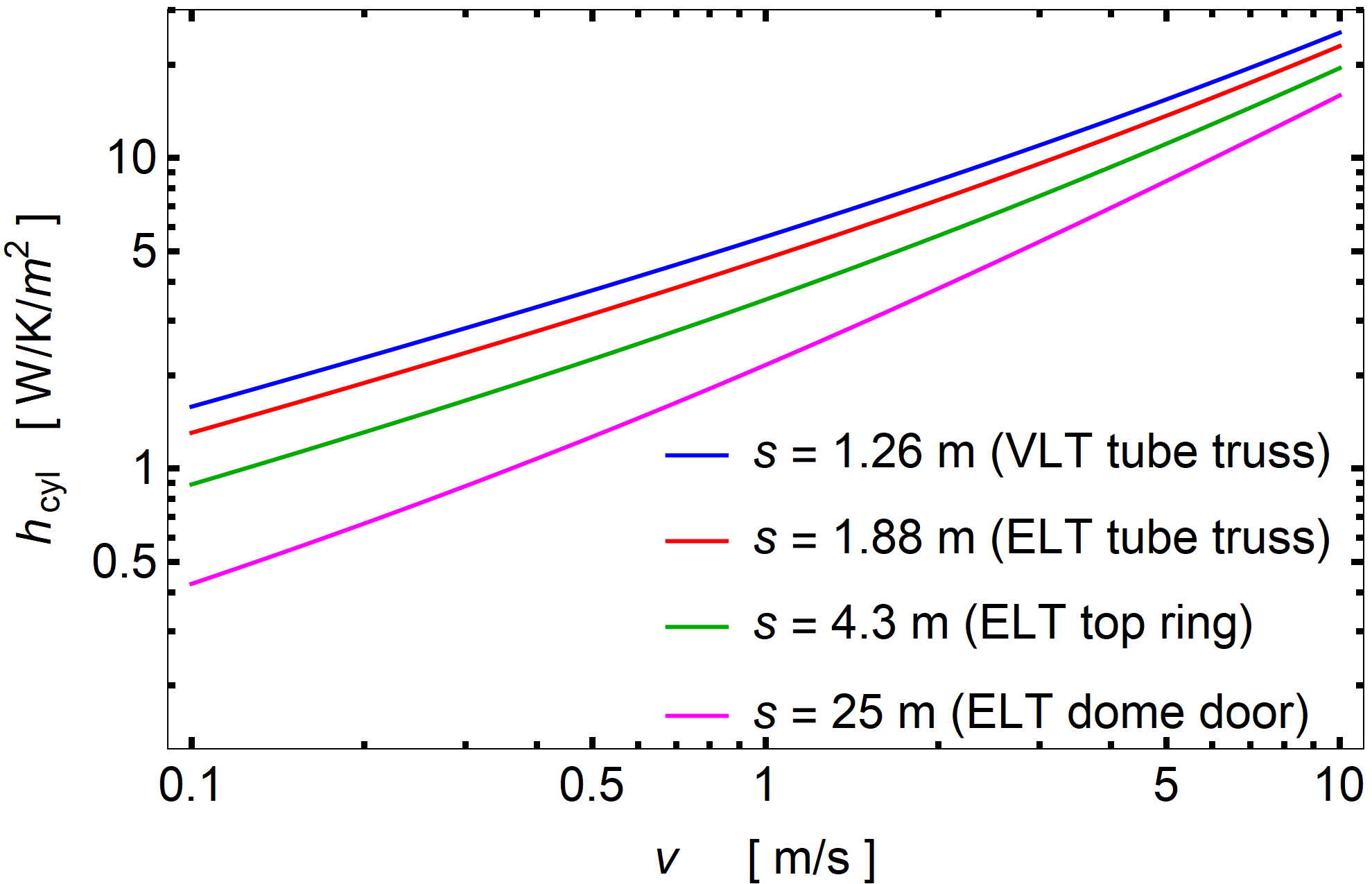}
}
\caption{Convective heat transfer coefficient~$h_\text{cyl}$ vs. local air speed~$v$ for three different interface lengths~$s$ representative of structures in the VLT and ELT telescopes, plus the ELT dome door.}
\label{fig:hcyl_vs_v}
\end{figure}

Figure\,\ref{fig:hcyl_vs_v} shows a plot of $h_\text{cyl}$ for three different values of~$s$ representative of structures in the VLT and ELT telescopes and the ELT dome door (all for the ELT environmental conditions and~$m=1.25\,\text{m}\,\text{W}^{-1}$).
\citet{Sunden2010} contains an overview of $h$ for cylinders with noncircular cross sections.

\begin{figure}[htb!]
\centerline{
\includegraphics[width=88mm]{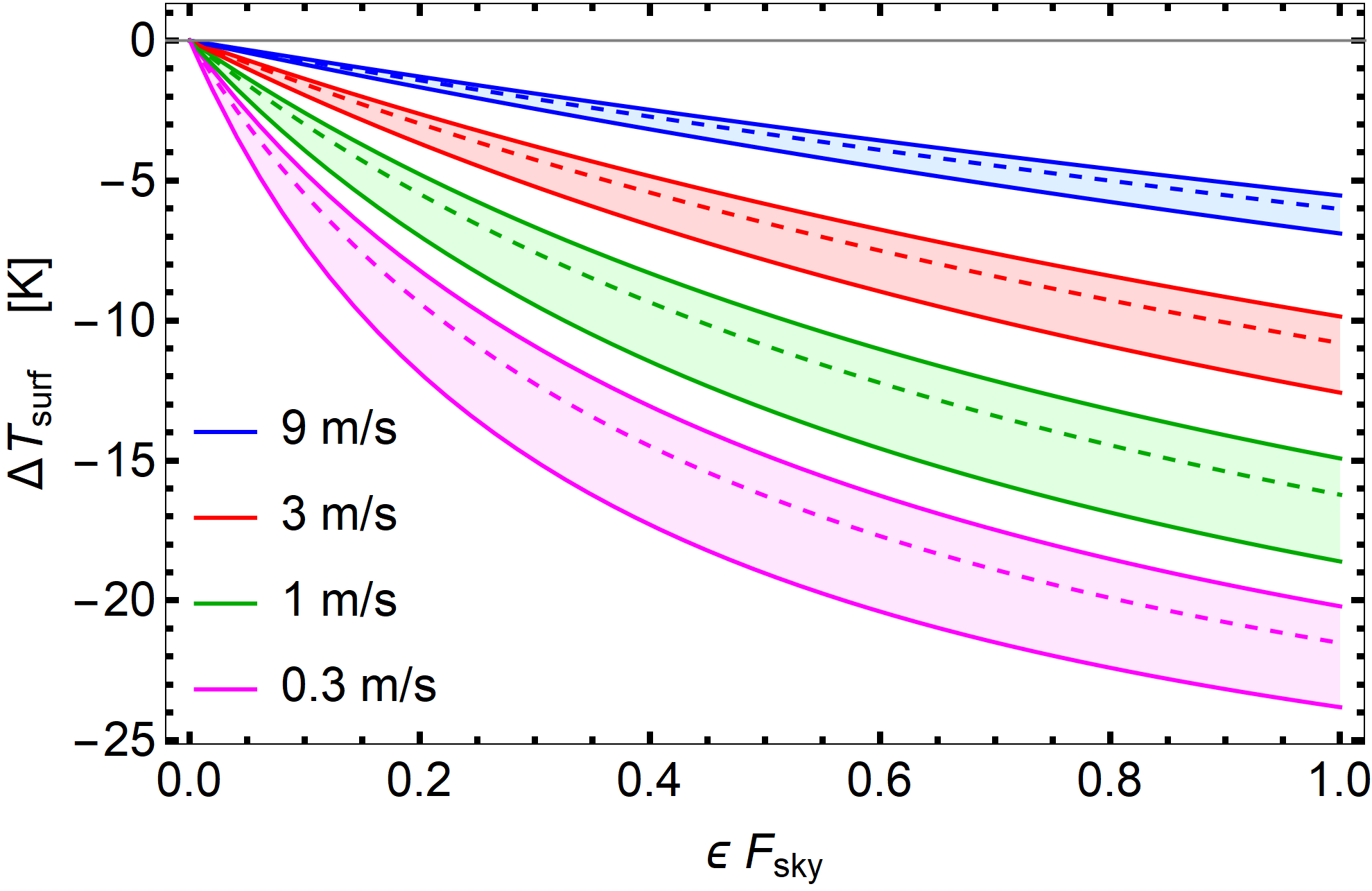}
}
\caption{Steady-state surface subcooling $\Delta T_\text{surf}$ as a function of $\overline{\epsilon}\,F_\text{sky}$ for four different local air speeds~$v$.\@ The upper edges of the colored fans correspond to $s = 1.26\,\text{m}$, the dashed lines to $s=1.88\,\text{m,}$ and the lower edges to $s = 4.3$\,m, respectively.}
\label{fig:DeltaTSurf_vs _EpsilonFSky}
\end{figure}

Figure\,\ref{fig:DeltaTSurf_vs _EpsilonFSky} shows a plot of $\Delta T_\text{surf}$ as a function of \,$\overline{\epsilon}\,F_\text{sky}$ according to Eq.\,(\ref{eq:DeltaTSurf3}), where the colors indicate four different local air speeds~$v$.\@ For these calculations and the remainder of this paper, we assume ELT median conditions as defined in Section~\ref{sec:meteo_ELT} for the zenith angle $\theta=37^\circ$.\@ The colored fans in Fig.\,\ref{fig:DeltaTSurf_vs _EpsilonFSky} are meant to give an impression of the influence of the interface length~$s$: As $s$ grows from $s=0.4\,\pi\,\text{m} = 1.26\,\text{m}$ to $s=4.3$\,m, $h_\text{cyl}$ diminishes and therefore $\eta$ grows, and thus the subcooling intensifies. Consequently, subcooling becomes more severe as telescopes grow.

\begin{figure}[htb!]
\centerline{
\includegraphics[width=88mm]{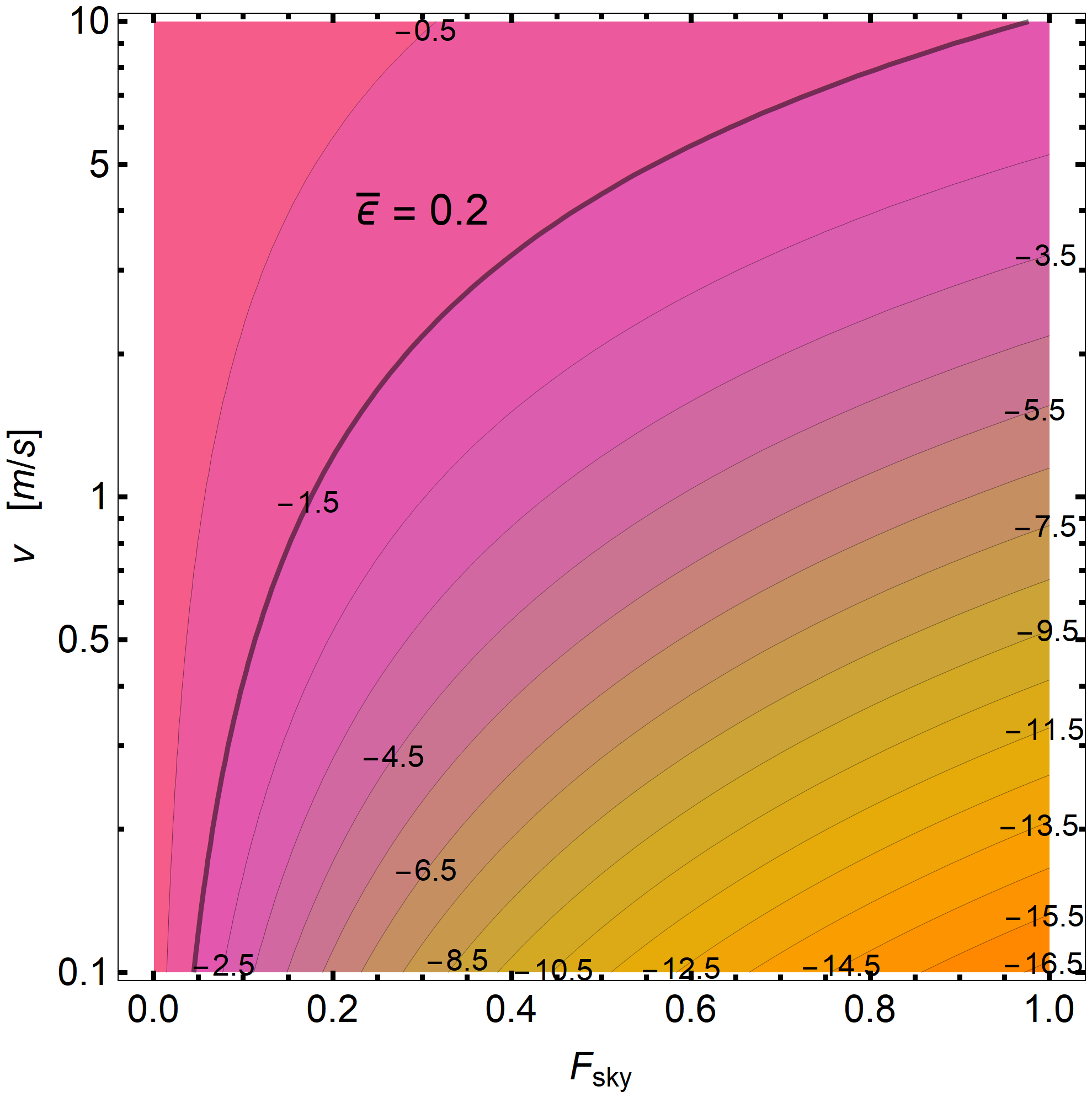}
}
\caption{Steady-state surface subcooling $\Delta T_\text{surf}$ as a function of $F_\text{sky}$ and $v$ for $s=4.3$\,m and~$\overline{\epsilon}=0.2$.\@ Contours:~Increments of $-1\,$K; thick contour:~Commonly accepted surface subcooling limit in telescopes of $\Delta T_\text{surf} = -1.5$\,K.}
\label{fig:DeltaTSurf_vs_FSky+v}
\end{figure}

Figure\,\ref{fig:DeltaTSurf_vs_FSky+v} is a two-dimensional plot of $T_\text{surf}(F_\text{sky},v)$ for $s=4.3$\,m (ELT top ring truss circumference) and $\overline{\epsilon}=0.2$, which is typical for dedicated low-emissivity paints; see~\citet{Yarbrough2010} for an overview of some respective commercial products. The tube structures of several large telescopes such as Gran Telescopio Canarias  (GTC, La~Palma), Keck~I+II (Mauna Kea, Hawaii), and the two Gemini telescopes (Cerro Pach{\'{o}}n, Chile, and Mauna~Kea) are covered with \emph{LO/MIT}
\footnote{https://web.archive.org/web/20170302132802/http://www.solec.org/ lomit-radiant-barrier-coating/lomit-technical-specifications/} 
paint by the company Solec (a silvery paint with small aluminum particles).
On surfaces that scatter visible stray light, one may apply selective foil such as {\it{Acktar NanoBlack}$^\text{\textregistered}$}\footnote{http://www.acktar.com/product/nano-black/} or {\it{Maxorb}} \citep{Mason1982}. These foils are black at visible wavelengths but highly reflective and thus low in emissivity ($\overline{\epsilon} < 10\%$) in the thermal infrared.\@
Combining Eq.\,(\ref{eq:h_cyl}) with Eq.\,(\ref{eq:DeltaTSurf3}), we find a simplified fit function near $\overline{\epsilon} = 0.2$ and $F_\text{sky} = 0.5$ given by
\begin{equation}
    \Delta T_\text{surf} \approx 0.636\, \Delta T_\text{sky}\;\overline{\epsilon}\, F_\text{sky}\,s^{\,0.355} \, v^{-0.551}, \label{eq:DeltaTSurf_Rule_of_Thumb}
\end{equation}
where $s$ is entered in meters and $v$ is in~meters per second, and again $\Delta T_\text{sky} = T_\text{sky} - T_\text{amb}$.\@ We found the power laws of $s$ and $v$ using the rule $b=x\,f'/f$ for a function $f(x) = a\,x^b$ with derivative~$f'$, applied at $s = 0.6\,\pi\,\text{m} = 1.88\,\text{m}$ and~$v = 1\,$m/s.
\subsection{Optical path difference}
Temperature differences $\Delta T$ in an air volume cause the optical path length difference (OPD) of
\begin{equation}
  \text{OPD} = \frac{\text{d}n}{\text{d}T} \int \Delta T\bigl(\vec{q}(l)\bigr) \; \text{d}l,
  \label{eq:OPD_T_integral}
\end{equation}
where $\vec{q}$ is a point in the air volume and $l$ measures geometrical distance along the ray path. The refractive index of air $n(\lambda,T,p)$ as a function of temperature $T$ and pressure $p$ is given by Edlen's modified formula~\citep{Boensch1998}.\@ Its derivative by $T$ can be written in simplified form as
\begin{equation}
   \frac{\text{d}n}{\text{d}T} = \frac{1}{\rho_0}\,\frac{\text{d}\rho}{\text{d}T} = -\,\frac{p}{\rho_0 \,R_\text{air}\, T^2},
    \label{eq:dndT}
\end{equation}
where $\rho = p/(R_\text{air}\,T)$ denotes the air density ($R_\text{air} = 287.06\,\text{J}\,\text{kg}^{-1}\,\text{K}^{-1}$ is the gas constant of dry air, yielding $\rho = 0.879\,\text{kg}\,\text{m}^{-3}$ at ELT median conditions) and the constant $\rho_0 = 4450\,\text{kg}\,\text{m}^{-3}$ in the wavelength range $0.4\,\mu\text{m} \leq \lambda \leq 3\,\mu\text{m}$.

At ELT median conditions, we obtain $\text{d}n/\text{d}T = -9.828 \times 10^{-12}\,\text{K}^{-1}\,\text{Pa}^{-1} \times p$ which is equal to $-7.0 \times 10^{-7}\,\text{K}^{-1}$.\@ For the ray path from the top of the ELT secondary mirror (M2) crown level to M1, then back to M2 and finally to the prime focus of total length $36\,\text{m}+31\,\text{m}+13\,\text{m} = 80\,\text{m}$, an air cooling as small as $\Delta T=25$\,mK causes an accumulated OPD of~$1.4\,\mu\text{m}$, equivalent to about $0.64\,\lambda$ in K-band ($2\,$--$\,2.4\,\mu$m).\@ The thermal OPD is thus significant and may actually become the leading wavefront error contribution in 30-meter-class  telescopes and greater.

\begin{figure}[ht!]
\centerline{
\includegraphics[width=88mm]{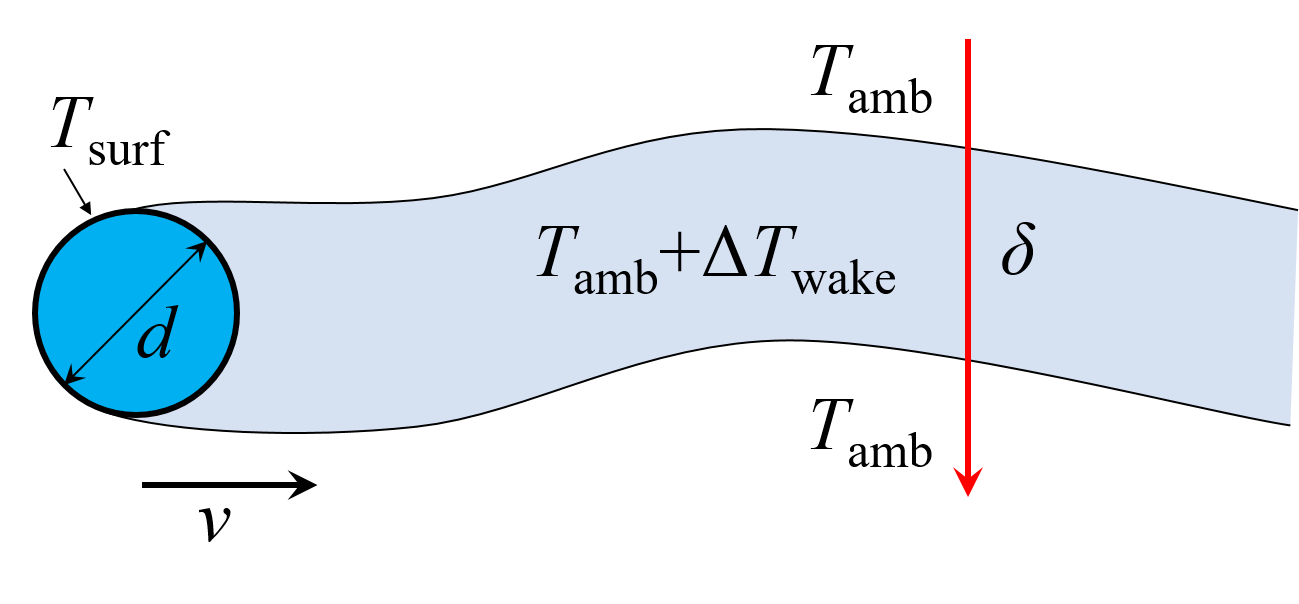}
}
\caption{Slice through a cylindrical truss of diameter $d = s/\pi$ running perpendicular to~$\vec{v}$ and to the telescope axis. Red arrow:~Ray of light traversing the cool wake of thickness $\delta$ and temperature difference~$\Delta T_\text{wake}$.}
\label{fig:wake}
\end{figure}

In the following, we establish the link between the subcooling temperature offset $\Delta T_\text{surf}$ and the OPD.\@ We rewrite the surface area in Eq.\,(\ref{eq:convective_heating}) as $A = s\,L$, where $L$ is the length of a cylindrical truss that we assume to be oriented perpendicular to the local airflow vector~$\vec{v}$ and perpendicular to the telescope optical axis, as would be the case for example for part of the top ring, as shown in Fig.\,\ref{fig:wake}.

On the downwind side of the truss, a wake of cooled air is generated. We consider now a ray of light in a thin air slice perpendicular to $\vec{v}$ and assume for simplicity that the wake has a thickness along the ray of $\delta$ with constant temperature offset~$\Delta T_\text{wake} < 0$.
The total volume of cooled air flowing away from the truss per time $\Delta t$ is then equal to $V/\Delta t = \delta\,L\,v$ and transports the heat flow
\begin{equation}
    -\dot{Q}_\text{conv} = \frac{V}{\Delta t}\, c_p\, \rho\, \Delta T_\text{wake}
    = c_p\, \rho\, v\,L\, \delta\,\Delta T_\text{wake},
    \label{eq:wake}
\end{equation}
where $c_p = 1005\;\text{J}\,\text{kg}^{-1}\,\text{K}^{-1}$ is the specific heat capacity of air at constant pressure. 

We can then derive the OPD imposed on the ray by the cooled air generated by a single truss of circumference $s$ in a single ray pass by combining Eq.\,(\ref{eq:wake}) with Eq.\,(\ref{eq:convective_heating})
\begin{equation}
\text{OPD} = \frac{\text{d}n}{\text{d}T}\,\delta\,\Delta T_\text{wake}
= \frac{\text{d}n}{\text{d}T}\, \frac{s\, h\, \Delta T_\text{surf}}{c_p\, \rho\, v}.
\label{eq:OPD_general}
\end{equation}
In reality, the wake is of course not isothermal, but instead the temperature distribution depends on the aerodynamics and is nonstationary. However, the actual temperature distribution along the ray is immaterial for the problem at hand since the path integral in Eq.\,(\ref{eq:OPD_T_integral}) is a conserved quantity equal to $\delta\,\Delta T_\text{wake}$ when averaged over time and along the truss length~$L$.\@ In the absence of this spatial and temporal averaging, transient vortexes can distribute air pockets of different temperature in the plane orthogonal to the rays \citep{MartinezConvection} generating a transient wavefront error; however, this aspect is beyond the scope of this work.

When $h_\text{rad} \ll h$ (e.g., with low-emissivity coatings and under most airflow conditions inside the dome) meaning that $\Delta T_\text{surf}$ is proportional to $\Delta T_D/h$, and as $\text{d}n/\text{d}T$ is proportional to $p$ and thus to $\rho$, we can simplify Eq.\,(\ref{eq:OPD_general}) to
\begin{equation}
  \text{OPD} \approx v_0\,\frac{\overline{\epsilon}\,F_\text{sky}\,s}{v}.
  \label{eq:OPD_simple}
\end{equation}
For ELT median conditions, $v_0 \approx 10^{-7}\,\text{m}/\text{s}$.\@ We point out that the OPD does not depend on~$h$: In steady state, the heat power lost by the radiation imbalance is exactly compensated by convective heat transfer from the inflowing air so that $h$ cancels.

Because the average air cooling is inversely proportional to the air mass passing the telescope structure per time, $\text{OPD} \propto v^{-1}$.\@ This fundamental relationship holds independently of the actual values of the structural subcooling $\Delta T_\text{surf}$ that various parts of the telescope converge to in steady state and explains the name "Low Wind Effect" \citep{Sauvage2015,Milli2018}; at very low wind speeds, forced convection gives way to natural convection so that the limit $v \rightarrow 0$ is never reached.

By the same argument, the OPD does not depend on air pressure either, and therefore telescopes at high altitude are in principle afflicted by the same magnitude of thermal OPD as at the ground level (more specifically, at high altitude the PWV column tends to be smaller, and therefore $\Delta T_\text{sky}$ grows in magnitude and $v_0$ increases).\@ 

The structures in a telescope causing large OPD are those with high-emissivity coatings (e.g., any regular industrial paint, irrespective of color) and those with large dimensions and/or high ratios~$F_\text{sky}/v$.\@ The sky view factor $F_\text{sky}$ tends to be larger for structures closer to the dome slit than those near the primary mirror, but the same holds for~$v$ \citep{Cullum2000}.\@ Moreover, we remind the reader that structures facing M1 may see a significant fraction of the sky in reflection. Therefore, it is not  necessarily clear which telescope structures are the most critical in terms of causing subcooling OPD.\

\begin{figure}[htb!]
\centerline{
\includegraphics[width=88mm]{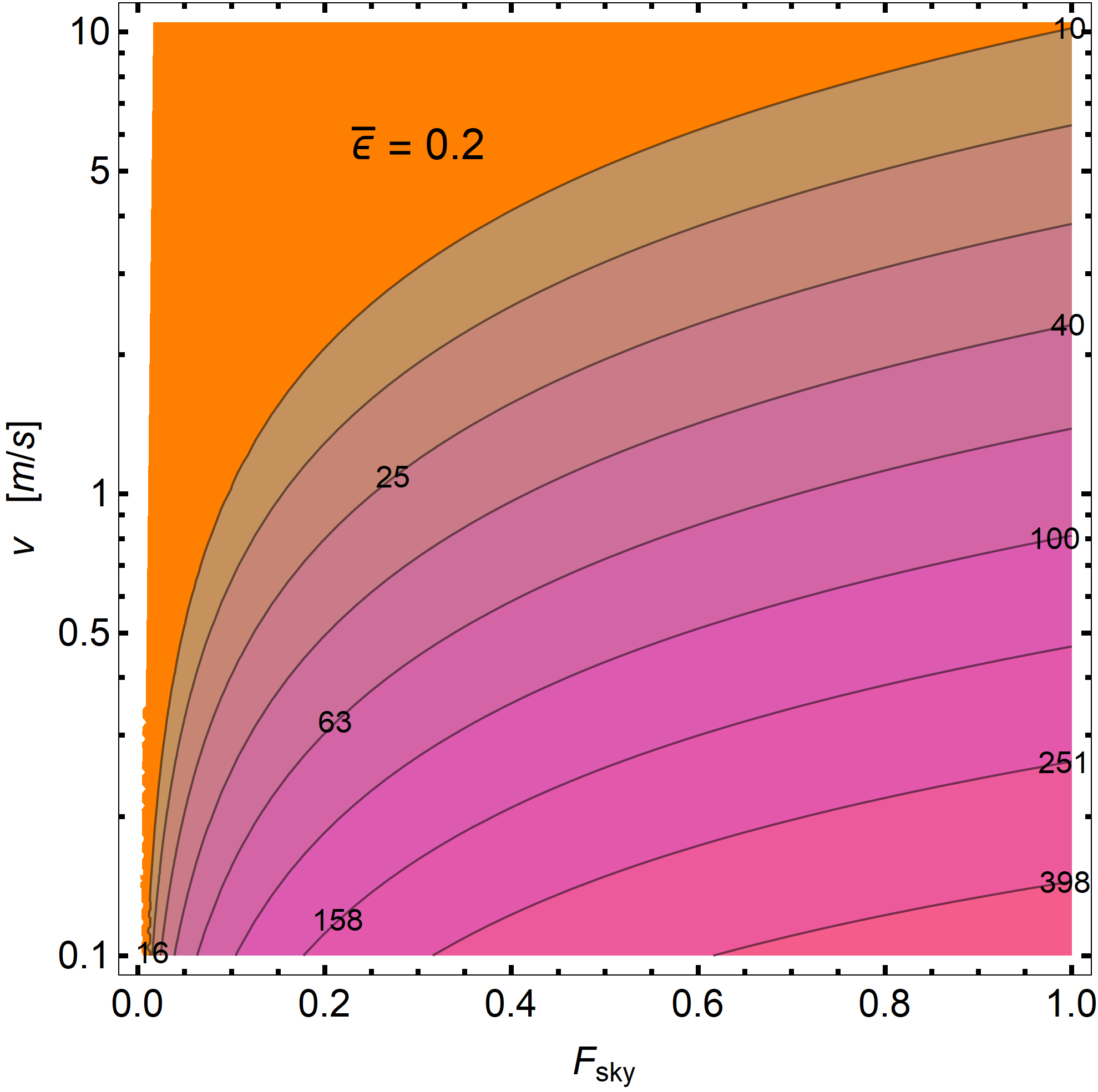}
}
\caption{Steady-state OPD from subcooling as a function of $F_\text{sky}$ and $v$ for a single ray pass through the wake of a single truss of circumference $s=4.3$\,m and~$\overline{\epsilon}=0.2$ (with ELT median conditions).\@ Contours:~OPD in nanometers in increments of factors of $10^{1/5}\approx 1.585$, starting at~$10\,$nm.}
\label{fig:OPD_vs_FSky+v}
\end{figure}

Figure\,\ref{fig:OPD_vs_FSky+v} shows a two-dimensional plot of the OPD as a function of $F_\text{sky}$ and $v$ for a single ray pass through the wake of a single truss of circumference $s=4.3$\,m (ELT top ring) and~$\overline{\epsilon}=0.2$, again at ELT median conditions.
The OPD from a single truss of typical size therefore amounts to only a few tens of nanometers. Moreover, the OPD caused by the top ring for example may tend to spread rather uniformly across the pupil, in which case it would not strongly distort the wavefront of starlight. However, the situation is different for vertically oriented structures such as vertical trusses in the telescope tube or a tertiary/quaternary mirror tower. Assuming that the cold wake thickness is similar to the truss diameter $\delta\approx d$, the OPD for structures running parallel to the optical axis must be scaled by a factor of up to the truss aspect ratio, for instance $L/d = 22/0.6 \approx 37$ for one of the vertical ELT telescope tube trusses.

To estimate how the OPD from subcooling scales with telescope size in general, we assume that telescopes of increasing sizes are supported by trusses whose circumference $s$ and length $L$ in the direction parallel to the telescope axis both scale linearly with the primary mirror diameter~$D$, causing the OPD from the wake of each truss to scale like$~D^2$.\@ This relationship holds for the structure above the M1 level only, as long as we assume a roughly identical focal ratio of the primary mirrors. However, the tube structure of the upcoming generation of 30-meter-class telescopes and greater  is becoming more delicate than that of the preceding 8- to 10-meter-class telescopes: The number of trusses increases, while each truss circumference grows only modestly. In the somewhat unrealistic limit that $s$ does not grow at all and $L$ rises proportionally to $D$, the OPD scales only linearly with~$D$ (but spreads more widely across the pupil).\@ In practice, the scaling power law will lie somewhere in between these extremes.

The rays from a star at center field propagating through a rotationally symmetric telescope travel in a plane defined by the entry point of the ray in the pupil and the optical axis. The closer the entry point is to the central obscuration (CO), the closer the ray segments between reflections are to each other. The air volume that in vertical projection is close to the rim of the CO is therefore traversed at least three times, boosting the OPD on the downwind side of the~CO.\@ Unfortunately, the M2 crown and the tertiary mirror tower introduce a lot of vertical surface area in this region with high~$F_\text{sky}$.\@ Moreover, any vertically extended tertiary mirror tower and the M2 crown are optically conjugated to a large range of heights above the telescope, which can induce field-dependent errors in the adaptive optics correction. A quantitative analysis requires extensive CFD simulations of the entire dome volume with temperatures \citep{Cho2011,Vogiatzis2014TMT,Ladd2016,Oser2018,Vogiatzis2018}.

\subsection{Heat conduction}
\label{subsec:conduction}

So far, we have neglected heat conduction within the structures, which will tend to equalize the structure temperature and thus spread the subcooling effect. In Appendix\,\ref{appendix:conduction_length_scale}, we derive the characteristic length scale due to heat conduction $\xi$ with typical values of a few tens of centimeters in telescope trusses (see Table\,\ref{table:material_properties}).

The OPD is approximately proportional to $\overline{\epsilon}\,F_\text{sky}\,s$ as Eq.\,(\ref{eq:OPD_simple}) shows. Therefore, it is fortunately equivalent to average the temperature around the perimeter of trusses when estimating the OPD in its wake, or\ for rectangular cross sections for example, to sum the OPD contributions of the four side faces. However, one should not assume that the subcooling spreads evenly along the length of a truss, which typically far exceeds~$\xi$.

\subsection{Transient effects}
\label{subsec:transient}

Telescope structures have a significant thermal inertia that delays temperature changes. In Appendix~\ref{appendix:thermal_inertia}, we derive the characteristic thermal relaxation timescale $\tau$ with typical values of a few hours (see Table\,\ref{table:material_properties}).

Sky subcooling is therefore a slow process that generally increases during the night. We note that any incorrect temperature conditioning of the telescope at sunset, including overheating ($\Delta T_\text{surf} > 0$), will generate additional OPD, starting already at the beginning of the observations \citep{Wilson2013}.

We point out that in contrast to the characteristic length $\xi$, the relaxation time $\tau $ is linearly dependent on~$w/h$, where $w$ is the truss wall thickness, and thus varies more strongly among different structures in a telescope, even if they are made of the same material. Additionally, the wind veers, the sky temperature drifts, there is variable cloud coverage, the telescope pointing zenith angle varies and so on, and therefore sky subcooling is an inherently transient process and any steady-state computation is only an approximation.

A serious source of wavefront errors in some telescopes is "mirror seeing" \citep{Wilson2013} caused by $\Delta T_\text{surf}$ on the optical mirror surface, which can sometimes become even more detrimental than "dome seeing" \citep{Racine1991}.\@ However, mirror seeing is not normally linked to radiative cooling because typical mirrors have broadband coatings, and therefore very low emissivity such as $\overline{\epsilon} < 2$\%.\@ However, subcooling can become a problem if the uncoated backside of a secondary mirror is (partly) exposed to the night sky (the emissivity of typical mirror substrates made of Zerodur$^\text{\textregistered}$ is $\approx 0.9$).

To compound the complexity, the ambient air temperature $T_\text{amb}$ also fluctuates. During a typical night, $T_\text{amb}$ drops quickly after sunset and decreases later more slowly, meaning that the radiation imbalance from the cold sky initially accelerates the thermalization of the telescope. Conversely, a rise in $T_\text{amb}$ exacerbates the subcooling. The median temporal gradients of $T_\text{amb}$ given in Section\,\ref{sec:meteo_ELT} are smaller than typical subcooling gradients of the order of $\dot{T}_\text{amb} \approx -1.5\,\text{K}/\tau \approx -1\,$K/h, but are not negligible.

One can shorten the subcooling relaxation time whilst lowering the emissivity and slowing down the structural deflection of the telescope structure by covering the truss surfaces with a thermal insulation layer topped by aluminum foil or thin sheet metal; a strategy chosen at the Subaru Telescope on Mauna~Kea, Hawaii \citep{Bely2006}.\@ However, we note that applying an insulation layer on top of a large thermal mass does not diminish the subcooling (i.e., does not mitigate $\Delta T_\text{surf}$ in the steady state) because the radiation imbalance on its surface is not reduced.


\section{Conclusions}
\label{sec:conc}
Subcooling of structures on the ground is caused by a thermal-radiation imbalance between the night sky and a structure on the ground at ambient temperature which can reach $150\,\text{W}\,\text{m}^{-2}$ and above. Structural elements in telescopes therefore cool by a few Kelvin (see Eqs.\,(\ref{eq:DeltaTSurf3} and \ref{eq:DeltaTSurf_Rule_of_Thumb})) and in turn cool the surrounding air. Wakes of cooled air flowing from elements in or near the telescope beam, such as spider trusses, the top ring, the tertiary mirror tower, and so on,\ cause an optical path length difference (OPD), introducing a transient wavefront error. While the size of modern telescopes continues to grow, the budgets of acceptable wavefront error are diminishing for diffraction-limited science cases. Therefore, subcooling poses a serious challenge to the generation of 30-meter telescopes and beyond, and second-generation instruments with adaptive-optics correction in existing 10-meter-class telescopes.

Additionally, we find the following:
\renewcommand{\labelitemi}{$\bullet$}
\begin{itemize}
\item The integrated thermal sky radiation can be characterized by the effective bolometric sky temperature $T_\text{sky}$.\@ This temperature is dominated by the lower atmosphere and $T_\text{sky}$ therefore floats with the ambient air temperature $T_\text{amb}$ near the ground. Based on spectra from the Cerro Paranal Sky Model, we derived a simple fit formula for $T_\text{sky}$ as a function of $T_\text{amb}$, the PWV column height, and the zenith angle $\theta$; see Eq.\,(\ref{eq:TSkyFit}).\@ The formula is valid for clear sky conditions, which are present at Cerro Paranal for about 90\% of the time, and can be applied to any relatively dry site. For the investigated range of PWV values from 0.5 to 20\,mm, the difference between $T_\text{sky}$ and $T_\text{amb}$ varies from $-50\,\text{K}$ to $-25\,\text{K}$ for $\theta = 0^\circ$.\@ The discrepancy can decrease by several Kelvin for higher zenith angles. 
\item
For Cerro Armazones, the site of the ELT, we estimate a median $T_\text{sky}$ of about 244\,K at a zenith angle of~$37^\circ$, which lies about $38\,$K below $T_\text{amb}$. 
%
\item The OPD due to subcooling of a single truss of circumference~$s$ is proportional to $\overline{\epsilon}\,F_\text{sky}\,s/v$, where $\overline{\epsilon}$ denotes surface emissivity, $F_\text{sky}$ is the sky view factor averaged around the truss cross section, and $v$ is the local air speed; see Eqs.\,(\ref{eq:OPD_general},\ref{eq:OPD_simple}).\@ As large telescopes are generally supported by trusses with both larger circumference and higher (vertical) length, the OPD scales between linearly and  quadratically with the primary mirror diameter.
\item It is crucial to apply low-emissivity coatings to all telescope structures above the primary mirror level that are exposed to a large fraction of the sky. Good choices are either bare aluminum or paints with embedded aluminum particles. On surfaces that may scatter visible stray light, one can apply selective foil that is black in the visible but reflective in the thermal infrared such as {\it{Acktar NanoBlack}$^\text{\textregistered}$} or {\it{Maxorb}}.\@ Conversely, regular industrial paint usually has high emissivity ($\overline{\epsilon} > 0.9$), even if it is white.
\item Sufficient airflow should be maintained in the telescope dome from sunset until the end of the night, for example\ by opening louvers. Because $\text{OPD} \propto \overline{\epsilon}\,F_\text{sky}/v$, the wavefront error from subcooling is not just caused by surfaces near the level of the dome slit, but also by any structures closer to the primary mirror if they are poorly ventilated and/or have high emissivity.
\item Special attention should be devoted to the outer dome cladding because of the combination of high sky view factor ($F_\text{sky} \approx 1$) and large surface area. It is advisable to use a low-emissivity coating, additionally with low solar absorptivity to limit daytime heating. Again, aluminum alloys or paints with aluminum pigments are good choices, while any colored paints should be avoided.
\item The most critical areas in a telescope structure in terms of wavefront distortion due to subcooling tend to be the vertical ones (i.e., running parallel to the optical axis), in particular surfaces near the projected rim of the central obscuration and the lateral spider truss faces. 
\item The ratio of heat capacity to convective efficiency of any telescope structure sets the timescale of thermal relaxation, but these quantities do not influence the steady-state subcooling temperature, as discussed in Section\,\ref{subsec:transient}.\@ The time to reach thermal equilibrium can be hours, and therefore subcooling often becomes more serious in the second half of the night.
%
\item Handheld commercial thermal cameras are not suitable to measure the bolometric sky temperature. These cameras normally use microbolometer detectors \citep{Rogalski2019}, which are only sensitive in the wavelength range of about $8\,$--$\,14\,\mu$m which largely overlaps with the astronomical N-band, a region with particularly high atmospheric transmission. On the other hand, if such cameras are employed to estimate subcooling of telescope structures, it is best to take the measurement at the end of the night directly after closing the dome to avoid measuring spuriously low temperatures due to partial sky reflection. 
\end{itemize}

\begin{acknowledgements} 
W.~Kausch is funded by the Hochschulraumstrukturmittel provided by the Austrian Federal Ministry of Education, Science and Research (BMBWF).\@ S.~Noll was financed by the project NO\,1328/1-1 of the German Research Foundation (DFG).\@ R.~Holzl{\"o}hner wishes to thank M.~Brinkmann (ESO) for stimulating discussions on fluid dynamics and A.~Ot{\'a}rola (TMT) for help with the Cerro Armazones environmental data statistics. The authors further thank J.~Vinther (ESO) for his continued support of~\emph{SkyCalc}.
\end{acknowledgements}

\bibliographystyle{aa}
\bibliography{references}

%
%
%
\begin{appendix}

\section{Heat conduction length scale}
\label{appendix:conduction_length_scale}
 We consider a plate and a pipe-shaped truss, both with wall thickness $w,$ letting $y$ be the local surface tangent coordinate along the truss length $L$ and $x$ run along the orthogonal tangential direction. We can write the two-dimensional heat equation for the temperature field $u(x,y) := \Delta T_\text{surf}(x,y)$ as
\begin{alignat}{2}
 & -k_m \left( u_{xx} + u_{yy}  \right) = q_\text{sky} - q_\text{surf} + q_\text{conv}
 \label{eq:conduction1}\\
 &\;\;\;\; = -\,\frac{h}{w} \left[ \bigl(1 + \eta\bigr)\,u - \eta\,\Delta T_D \right],\nonumber
\end{alignat}
%
where $u_{xx}$, $u_{yy}$ denote the respective second partial derivatives of $u$ by $x$ and $y$, $q = \dot{Q}/(A\,w)$ is the specific lateral heat flow per bulk wall volume (unit $\text{W}\,\text{m}^{-3}$) and we have inserted the heat flow terms on the left-hand side of Eq.\,(\ref{eq:Q_balance}).\@ In Eq.\,(\ref{eq:conduction1}), we use Eq.\,(\ref{eq:h_rad}) where now $h_\text{rad} = h_\text{rad}(x,y)$, $k_m$ is the thermal conductivity of the truss wall material and we assume that the heat conduction normal to the surface is so efficient that the wall becomes isothermal along $z$, thus $u(z) = \text{const}$ and~$k_m/w \gg h$.

We define the characteristic length $\xi$ which quantifies the ratio of lateral heat conduction in the truss wall to convection and rewrite Eq.\,(\ref{eq:conduction1}) as
\begin{alignat}{2}
 &\xi^2 \left( u_{xx} + u_{yy} \right) - u(x,y) + U_\text{rad}(x,y) = 0, \label{eq:conduction2}\\
 &\xi := \sqrt{\frac{w\, k_m}{h}},
\end{alignat}
where we assume again $h_\text{rad} \ll h$ as discussed after Eq.\,(\ref{eq:h_rad}) and the radiative forcing term $U_\text{rad}(x,y)$ is equal to $\eta\,\Delta T_D$.

We now consider two different simple structures: The first is a rectangular plate of dimensions $L \times s \times w$ with $L \gg s \gg w$.\@ We require that there be no heat flux across the plate edges, hence the temperature becomes stationary there. Further, the radiative forcing term contains a step function centered at $x=0$, for example because the plate is bent into an L-shaped form with a $90^\circ$ corner running parallel to the long axis $y$ where the sky view factor changes abruptly:
\begin{alignat}{2}
   u_x(\pm s/2) &= 0,\nonumber\\
   U_\text{rad}(x,y) &:= \Delta U\,\text{sgn}(x) + U_0,
  \label{eq:BC Plate}
\end{alignat}
where $u_x$ is the partial derivative of $u$ by $x$ and $\text{sgn}(x) = 1$ for $x > 0$, $\text{sgn}(x) = -1$ for $x < 0$ and $0$ otherwise. The solution of Eq.\,(\ref{eq:conduction2}) with the boundary conditions in Eq.\,(\ref{eq:BC Plate}) is
\begin{equation}
  u(x') = \Delta U \left[\tanh\bigl(s'\bigr)\sinh\bigl(x'\bigr) - \text{sgn}(x) \Bigl( \cosh\bigl(x'\bigr) - 1 \Bigr)\right] + U_0,\label{eq:u(x) Plate}
\end{equation}
where $\tanh$, $\sinh,$ and $\cosh$ denote the hyperbolic equivalents of the $\tan$, $\sin,$ and $\cos$ functions, respectively, and we have introduced the reduced variables $x' := x/\xi$ and~$s' := s/(2\xi)$.
\begin{figure}[htb!]
\centerline{
\includegraphics[width=88mm]{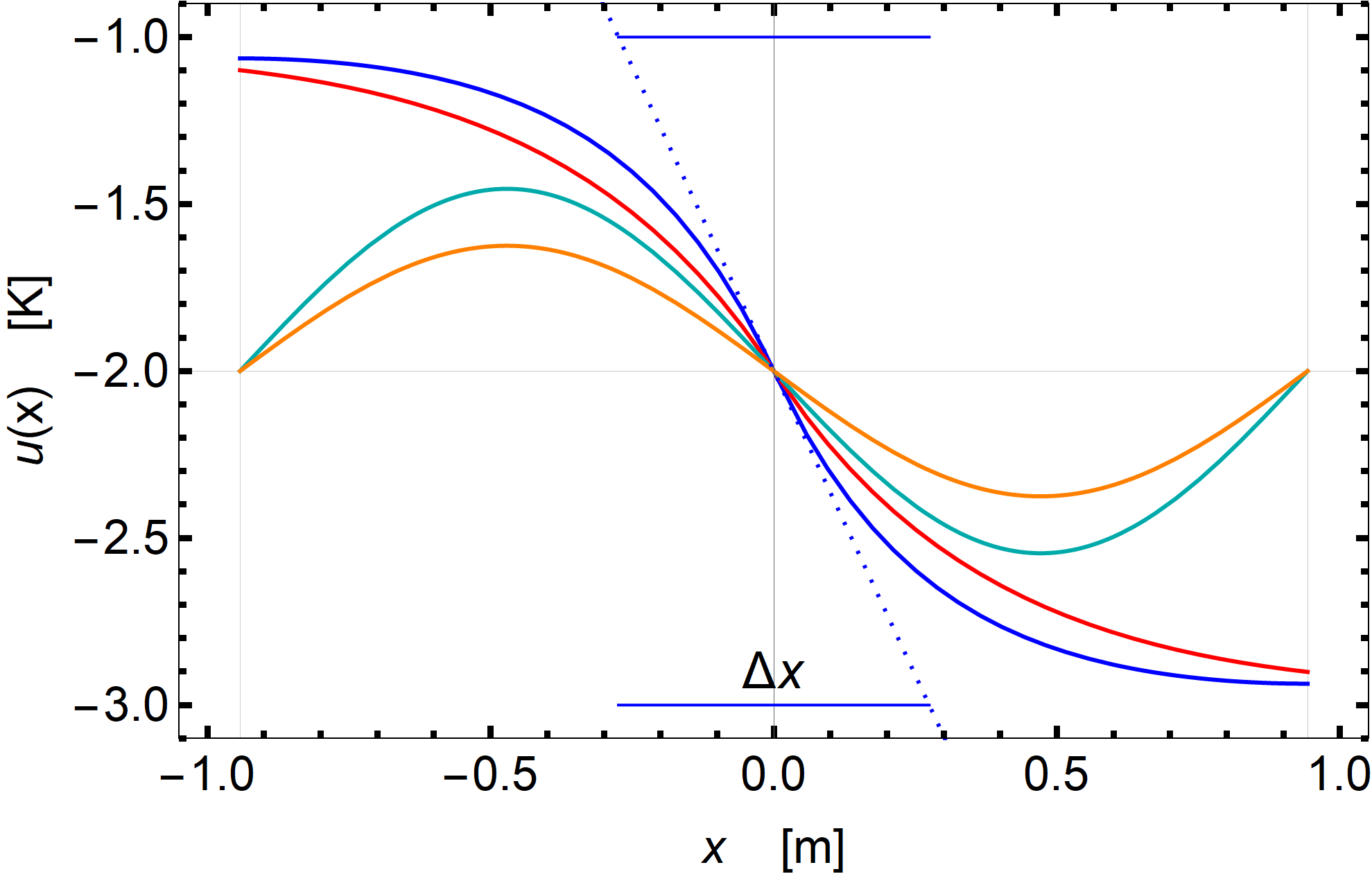}
}
\caption{Example of steady-state temperature distribution $u(x)$ with heat conduction. Blue and red curves: Plate of size $\pi\,0.6$\,m with step-like radiative forcing; $\xi = 0.27\,$m and $0.39$\,m, respectively. Turquoise and orange: Same $\xi$ values, but sinusoidal forcing around pipe-shaped truss of diameter $d = s/\pi = 0.6\,$m.\@ The transition region of the blue curve has the width $\Delta x = 0.55\,$m.}
\label{fig:u(x) conduction}
\end{figure}

\begin{table}
\caption{Properties of some materials commonly used in telescopes with derived characteristic lengths $\xi$ and relaxation times $\tau$, all for $w = 10$\,mm and $h = 4.5\,\text{W}\,\text{K}^{-1}\,\text{m}^{-2}$}             
\label{table:material_properties}      
\small
\begin{tabular}{c c  c c c c}        
\hline\hline                 
\noalign{\smallskip}
Quan-     &  Unit & Steel & Stainless  & Al   & Schott \\
tity      &       & S355\tablefootmark{a}  & steel 304$^b$  & 5754\tablefootmark{b} & Zerodur\tablefootmark{c,d} \\    
\noalign{\smallskip}
\hline                        
\noalign{\smallskip}
$k_m$ & $\text{W}\,\text{K}^{-1}\,\text{m}^{-1}$ & 45 & 16 & 125 & 1.46 \\
$\rho_m$ & $\text{kg}\,\text{m}^{-3}$ & 7850 & 8020 & 2670 & 2530 \\
$c_m$ & $\text{J}\,\text{K}^{-1}\,\text{kg}^{-1}$ & 450 & 480 & 900 & 821 \\
$\overline{\epsilon}$ & \% & n/a & 10--45  & 6--20 & 90 \\
\noalign{\smallskip}
\hline
\noalign{\smallskip}
$\xi$  & m     & 0.32 & 0.19 & 0.53 & 0.06 \\
$\tau$ & hours & 2.18 & 2.38 & 1.48 & 1.28 \\
\noalign{\smallskip}
\hline 
\noalign{\smallskip}
\end{tabular}
\tablefoottext{a}{{http://www.matweb.com}}\protect{\newline}
\tablefoottext{b}{\citet{Valencia2010}}\protect{\newline}
\tablefoottext{c}{{https://www.schott.com/en-gb/products/zerodur/technical-details}}\protect{\newline}
\tablefoottext{d}{\citet{Banyal2011}}

\end{table}

The blue and red curves in Fig.\,\ref{fig:u(x) conduction} show~$u(x)$, where $s=\pi\,0.6\,\text{m} \approx 1.88$\,m, which has the shape of an inverted letter~S with its inflection point at~$(x\!=\!0,\,u\!=\!U_0)$.\@ For this plot we used the material properties of construction steel S355 as listed in Table\,\ref{table:material_properties} and chose $w = 10$\,mm, $\Delta U = -1\,\text{K,}$ and $U_0 = -2\,\text{K}$.\@ We assumed the two different conductive heat transfer coefficients $h_1 = 6\,\text{W}\,\text{K}^{-1}\,\text{m}^{-2}$ and $h_2 = 3\,\text{W}\,\text{K}^{-1}\,\text{m}^{-2}$, corresponding to the scaling lengths $\xi_1 = 0.27$\,m and $\xi_2 = 0.39$\,m (blue and red curves, respectively).

We can define a transition region width $\Delta x$ satisfying $u_x(0)\,\Delta x/ 2 = \Delta U$, as indicated by the blue dotted line in Fig.\,\ref{fig:u(x) conduction}. Using $u_x(0) = \Delta U\, \tanh(s')/\xi$, we find
\begin{equation}
  \frac{\Delta x}{s} = \frac{1}{s'\,\tanh(s')} \approx 0.33 - 0.11\, \bigl(s' - 1\bigr) + O\,\bigl(s'^2 \bigr),
  \label{eq:Delta x}
\end{equation}
where the last term is the Taylor expansion about the value~$s'= 3$ relevant for the presented examples. We conclude that the transition region can cover a significant fraction of a typical steel plate.

The second structure is a pipe-shaped truss of diameter~$d = s/\pi$ with sinusoidal radiative forcing around its circumference instead of the step-like forcing, which could be a model of a vertical truss in the telescope tube, where $F_\text{sky}$ peaks on the side facing the telescope axis and becomes minimal on the opposite side oriented towards the inside of the dome. We modify the plate boundary conditions so that the problem becomes periodic in the circumferential coordinate~$x$, as already analyzed by Joseph \citet{Fourier1822}
\begin{alignat}{2}
   u(s/2) &= u(-s/2),\nonumber\\
   u_x(s/2) &= 0,\nonumber\\
   U_\text{rad}(x,y) &= \Delta U \cos\bigl(2\pi\,j\,x/s\bigr) + U_0,
  \label{eq:BC Pipe}
\end{alignat}
where $j$ is any integer. As in the plate problem, $U_\text{rad}(x,y)$ lies between the extremes $U_0 + \Delta U$ and $U_0 - \Delta U$, which are assumed now at $x=0$ and $x = \pm s/2$, respectively. The solution is
\begin{alignat}{2}
  u(x') &= \Psi_j\,\Delta U \cos\bigl(\pi\,j\,x'/s'\bigr) + U_0,\label{eq:u(x) Pipe}\\
  \Psi_j &:= \frac{{s'}^2}{{s'}^2+\pi^2\,j^2} = \frac{1}{1+\bigl(2\pi\,j\,\xi/s \bigr)^2},
  \label{eq:u(x)_Pipe}
\end{alignat}
which is plotted in Fig.\,\ref{fig:u(x) conduction} in the turquoise and orange curves for $j = 1$ and $s = 0.6\,\pi\,\text{m} = 1.88\,\text{m}$ and again setting $\xi_1 = 0.27$\,m and $\xi_2 = 0.39$\,m, respectively (as $u(x')$ is periodic in the pipe case, we have chosen to shift it along $x$ so that its inflection point lies at $x=0$ in the plot, as for the plate solution).\@ The factor $\Psi_j$ compresses the amplitude of the sinusoidal temperature variation around the pipe, in particular if~$s < 2\pi\,j\,\xi$.\@ We note that we can approximate any periodic boundary condition by a Fourier series, hence the linear combination of solutions in Eq.\,(\ref{eq:u(x)_Pipe}) with different spatial frequencies~$j$.

If a plate is thermally insulated on one side, such as can be the case for the outer dome cladding, the convective heat exchange on the insulated side is almost turned off. Conversely, for plates whose two side faces are exposed to the ambient air, the interface surface area~$A$ is twice that of the insulated case and $F_\text{sky}$ should be averaged over the two sides. However, in such a case the average $F_\text{sky}$ cannot exceed a value of one-half because of geometric constraints, and therefore the product $F_\text{sky}\,A$ may in some cases not differ strongly from the insulated plate case.

Inside a sealed air-filled structure such as a pipe truss, the enclosed air exchanges heat energy with the internal walls and there is also internal thermal radiation. However, as the internal airflow is usually not forced and because the wall temperature variation is normally too small to induce a significant radiation imbalance, we can neglect these factors.

Summarizing the plate and the pipe examples, we find that $u(x)$ becomes compressed if the characteristic length $\xi = (w\,k_m/h)^{1/2}$ is large compared to~$s$, that is, if lateral conduction dominates over convection. Irrespective of this condition, the average of $u(x)$ around the truss circumference is always equal to the average of $U_\text{rad}$, thus $U_0 = -2\,$K in our two examples. The limit $s \ll 2\pi\,\xi$ may be referred to as the isothermal regime, while the opposite limit $s \gg 2\pi\,\xi$ may be thought of as thermally disconnected. For a typical telescope pipe steel truss with a wall thickness of $w=10$\,mm, neither limit clearly applies around its circumference. However, the thermally disconnected regime applies along a typical length $L \gg 2\pi\,\xi$.\@
\section{Thermal inertia}
\label{appendix:thermal_inertia}
Until now, we have discussed only the steady-state solution. We now finally consider the full time-dependent heat equation for a three-dimensional body in air with temperature offset $u = u(x,y,z,t) := T_\text{body}(x,y,z,t) - T_\text{amb}(t)$ with time $t,$
\begin{equation}
 c_m\,\rho_m \left( u_t + \dot{T}_\text{amb} \right) - k_m \left( u_{xx} + u_{yy} + u_{zz}  \right)
  = q_\text{sky} - q_\text{surf} + q_\text{conv},
  \label{eq:conduction+transient}
\end{equation}
%
where $c_m$ and $\rho_m$ are the specific heat capacity and the density of the body material, respectively.

We can consider the pipe problem of the previous section with $j=1$ and the initial condition $u(x,y,z,0) = 0$ and, setting $u_{zz} = \dot{T}_\text{amb} = 0$, find the solution
\begin{equation}
 u(x',t') = \Psi_1 \,\Delta U\, \cos \left(\pi\,x'/s'\right) \Bigl(1-e^{-t'/\Psi_1}\Bigr) + U_0 \Bigl(1-e^{-t'}\Bigr)
\end{equation}
as a function of the reduced time~$t'= t/\tau$ with the lateral thermal relaxation time
\begin{equation}
 \tau := \frac{w\,c_m\,\rho_m}{h}.
\end{equation}
We now evaluate the time $t = \tau$ after which the term $(1-\exp(-t/\tau))$  grows from zero to $1 - 1/e \approx 0.63$.\@ We assume again the material properties of construction steel S355 listed in Table\,\ref{table:material_properties} and $w = 10$\,mm to arrive at $\tau_1 = 1.64$\,hr and $\tau_2 = 3.27$\,hr for $h_1 = 6\,\text{W}\,\text{K}^{-1}\,\text{m}^{-2}$ and $h_2 = 3\,\text{W}\,\text{K}^{-1}\,\text{m}^{-2}$, respectively.

Table\,\ref{table:material_properties} shows an overview of some material properties applicable to telescopes. We do not provide an emissivity value for S355 steel since it depends strongly on the surface preparation and regular steel is normally coated anyway to protect from rust. The aluminum alloy 5754 is a candidate for the ELT outer dome cladding ($\overline{\epsilon} = 6$\% corresponds to a clean and 20\% to a contaminated surface, respectively).\@

Zerodur is a low-thermal-expansion glass ceramic used for various telescope mirrors. The thermal conductivity of Zerodur and other glasses and ceramics is often so low that the relation $k_m/w \gg h$ no longer holds for mirrors with typical thickness values of $w=50\,$mm (ELT M1 segments) to hundreds of millimeters. As a consequence, transient thermal phenomena become three-dimensional problems \citep{Banyal2011} with the relaxation time along $z$ of $\tau_z = w^2\,c_m\,\rho_m/k_m$.

\end{appendix}
%


\end{document}